\pdfoutput=1

\documentclass[12pt,a4paper]{article}

\usepackage{setspace}

\usepackage{ifthen} 
\newboolean{pdflatex}
\setboolean{pdflatex}{true} 

\newboolean{articletitles}
\setboolean{articletitles}{true} 

\newboolean{uprightparticles}
\setboolean{uprightparticles}{false} 

\newboolean{inbibliography}
\setboolean{inbibliography}{false} 

\usepackage[top=1in, bottom=1.25in, left=1in, right=1in]{geometry}

\usepackage{microtype}
\usepackage{lineno}  
\usepackage{xspace} 
\usepackage{caption} 

\usepackage{graphicx}  
\usepackage{color}
\usepackage{colortbl}
\graphicspath{{./figs/}} 

\usepackage{amsmath} 
\usepackage{amssymb}
\usepackage{amsfonts}
\usepackage{upgreek} 

\newcommand*\patchAmsMathEnvironmentForLineno[1]{%
\expandafter\let\csname old#1\expandafter\endcsname\csname #1\endcsname
\expandafter\let\csname oldend#1\expandafter\endcsname\csname
end#1\endcsname
 \renewenvironment{#1}%
   {\linenomath\csname old#1\endcsname}%
   {\csname oldend#1\endcsname\endlinenomath}%
}
\newcommand*\patchBothAmsMathEnvironmentsForLineno[1]{%
  \patchAmsMathEnvironmentForLineno{#1}%
  \patchAmsMathEnvironmentForLineno{#1*}%
}
\AtBeginDocument{%
\patchBothAmsMathEnvironmentsForLineno{equation}%
\patchBothAmsMathEnvironmentsForLineno{align}%
\patchBothAmsMathEnvironmentsForLineno{flalign}%
\patchBothAmsMathEnvironmentsForLineno{alignat}%
\patchBothAmsMathEnvironmentsForLineno{gather}%
\patchBothAmsMathEnvironmentsForLineno{multline}%
\patchBothAmsMathEnvironmentsForLineno{eqnarray}%
}

\usepackage{hyperref}    
\usepackage[all]{hypcap} 

\def\lhcb {\mbox{LHCb}\xspace}

\def\velo   {VELO\xspace}

\def\MagUp {\mbox{\em Mag\kern -0.05em Up}\xspace}

\ifthenelse{\boolean{uprightparticles}}%
{

 \def\Pmu         {\ensuremath{\upmu}\xspace}

 \def\Ppsi        {\ensuremath{\uppsi}\xspace}

 \def\PDelta      {\ensuremath{\Delta}\xspace}                 
 \def\PXi      {\ensuremath{\Xi}\xspace}                 
 \def\PLambda      {\ensuremath{\Lambda}\xspace}                 
 \def\PSigma      {\ensuremath{\Sigma}\xspace}                 
 \def\POmega      {\ensuremath{\Omega}\xspace}                 
 \def\PUpsilon      {\ensuremath{\Upsilon}\xspace}                 
 
 \def\PB      {\ensuremath{\mathrm{B}}\xspace}                 
                  
 \def\PD      {\ensuremath{\mathrm{D}}\xspace}

 \def\PH      {\ensuremath{\mathrm{H}}\xspace}                 
                  
 \def\PJ      {\ensuremath{\mathrm{J}}\xspace}                 
 \def\PK      {\ensuremath{\mathrm{K}}\xspace}

 \def\Pb      {\ensuremath{\mathrm{b}}\xspace}                 
 \def\Pc      {\ensuremath{\mathrm{c}}\xspace}

 \def\Pi      {\ensuremath{\mathrm{i}}\xspace}

 \def\Pt      {\ensuremath{\mathrm{t}}\xspace}

}
{

 \def\Pmu         {\ensuremath{\mu}\xspace}

 \def\Ppsi        {\ensuremath{\psi}\xspace}                 
                  
 \mathchardef\PDelta="7101
 \mathchardef\PXi="7104
 \mathchardef\PLambda="7103
 \mathchardef\PSigma="7106
 \mathchardef\POmega="710A
 \mathchardef\PUpsilon="7107
                  
 \def\PB      {\ensuremath{B}\xspace}                 
                  
 \def\PD      {\ensuremath{D}\xspace}

 \def\PH      {\ensuremath{H}\xspace}                 
                  
 \def\PJ      {\ensuremath{J}\xspace}                 
 \def\PK      {\ensuremath{K}\xspace}

 \def\Pb      {\ensuremath{b}\xspace}                 
 \def\Pc      {\ensuremath{c}\xspace}

 \def\Pi      {\ensuremath{i}\xspace}

 \def\Pt      {\ensuremath{t}\xspace}

}

\makeatletter
\ifcase \@ptsize \relax
  \newcommand{\miniscule}{\@setfontsize\miniscule{4}{5}}
\or
  \newcommand{\miniscule}{\@setfontsize\miniscule{5}{6}}
\or
  \newcommand{\miniscule}{\@setfontsize\miniscule{5}{6}}
\fi
\makeatother

\DeclareRobustCommand{\optbar}[1]{\shortstack{{\miniscule (\rule[.5ex]{1.25em}{.18mm})}
  \\ [-.7ex] $#1$}}

\def\mumu       {{\ensuremath{\Pmu^+\Pmu^-}}\xspace}

\def\H      {{\ensuremath{\PH^0}}\xspace}

\def\cquark    {{\ensuremath{\Pc}}\xspace}
\def\cquarkbar {{\ensuremath{\overline \cquark}}\xspace}
\def\ccbar     {{\ensuremath{\cquark\cquarkbar}}\xspace}
\def\bquark    {{\ensuremath{\Pb}}\xspace}
\def\bquarkbar {{\ensuremath{\overline \bquark}}\xspace}
\def\bbbar     {{\ensuremath{\bquark\bquarkbar}}\xspace}
\def\tquark    {{\ensuremath{\Pt}}\xspace}
\def\tquarkbar {{\ensuremath{\overline \tquark}}\xspace}
\def\ttbar     {{\ensuremath{\tquark\tquarkbar}}\xspace}

\def\kaon    {{\ensuremath{\PK}}\xspace}
  \def\Kbar    {{\kern 0.2em\overline{\kern -0.2em \PK}{}}\xspace}

\def\KorKbar    {\kern 0.18em\optbar{\kern -0.18em K}{}\xspace}

\def\Kstarz  {{\ensuremath{\kaon^{*0}}}\xspace}

  \def\Dbar    {{\kern 0.2em\overline{\kern -0.2em \PD}{}}\xspace}

\def\DorDbar    {\kern 0.18em\optbar{\kern -0.18em D}{}\xspace}

\def\Bbar    {{\ensuremath{\kern 0.18em\overline{\kern -0.18em \PB}{}}}\xspace}

\def\BorBbar    {\kern 0.18em\optbar{\kern -0.18em B}{}\xspace}

\def\jpsi     {{\ensuremath{{\PJ\mskip -3mu/\mskip -2mu\Ppsi\mskip 2mu}}}\xspace}

  \def\Y#1S{\ensuremath{\PUpsilon{(#1S)}}\xspace}

\def\Lbar        {{\ensuremath{\kern 0.1em\overline{\kern -0.1em\PLambda}}}\xspace}
\def\LorLbar    {\kern 0.18em\optbar{\kern -0.18em \PLambda}{}\xspace}

\newcommand{\decay}[2]{\ensuremath{#1\!\to #2}\xspace}         

\def\to                 {\ensuremath{\rightarrow}\xspace}

\def\AT#1     {\ensuremath{A_{\mathrm{T}}^{#1}}\xspace}           

\def\C#1      {\ensuremath{\mathcal{C}_{#1}}\xspace}                       
\def\Cp#1     {\ensuremath{\mathcal{C}_{#1}^{'}}\xspace}                    
\def\Ceff#1   {\ensuremath{\mathcal{C}_{#1}^{\mathrm{(eff)}}}\xspace}        
\def\Cpeff#1  {\ensuremath{\mathcal{C}_{#1}^{'\mathrm{(eff)}}}\xspace}       
\def\Ope#1    {\ensuremath{\mathcal{O}_{#1}}\xspace}                       
\def\Opep#1   {\ensuremath{\mathcal{O}_{#1}^{'}}\xspace}                    

\newcommand{\tev}{\ifthenelse{\boolean{inbibliography}}{\ensuremath{~T\kern -0.05em eV}\xspace}{\ensuremath{\mathrm{\,Te\kern -0.1em V}}}\xspace}
\newcommand{\gev}{\ensuremath{\mathrm{\,Ge\kern -0.1em V}}\xspace}
\newcommand{\mev}{\ensuremath{\mathrm{\,Me\kern -0.1em V}}\xspace}
\newcommand{\kev}{\ensuremath{\mathrm{\,ke\kern -0.1em V}}\xspace}
\newcommand{\ev}{\ensuremath{\mathrm{\,e\kern -0.1em V}}\xspace}
\newcommand{\gevc}{\ensuremath{{\mathrm{\,Ge\kern -0.1em V\!/}c}}\xspace}
\newcommand{\mevc}{\ensuremath{{\mathrm{\,Me\kern -0.1em V\!/}c}}\xspace}
\newcommand{\gevcc}{\ensuremath{{\mathrm{\,Ge\kern -0.1em V\!/}c^2}}\xspace}
\newcommand{\gevgevcccc}{\ensuremath{{\mathrm{\,Ge\kern -0.1em V^2\!/}c^4}}\xspace}
\newcommand{\mevcc}{\ensuremath{{\mathrm{\,Me\kern -0.1em V\!/}c^2}}\xspace}

\def\m    {\ensuremath{\rm \,m}\xspace}

\def\cm   {\ensuremath{\rm \,cm}\xspace}

\def\mm   {\ensuremath{\rm \,mm}\xspace}

\def\mum  {\ensuremath{{\,\upmu\rm m}}\xspace}

\def\pb {\ensuremath{\rm \,pb}\xspace}
\def\invpb {\ensuremath{\mbox{\,pb}^{-1}}\xspace}

\def\invfb   {\ensuremath{\mbox{\,fb}^{-1}}\xspace}

\def\ps   {\ensuremath{{\rm \,ps}}\xspace}

\newcommand{\chisqndf}{\ensuremath{\chi^2/\mathrm{ndf}}\xspace}

\def\gsim{{~\raise.15em\hbox{$>$}\kern-.85em
          \lower.35em\hbox{$\sim$}~}\xspace}
\def\lsim{{~\raise.15em\hbox{$<$}\kern-.85em
          \lower.35em\hbox{$\sim$}~}\xspace}

\def\sqs   {\ensuremath{\protect\sqrt{s}}\xspace}

\def\ptot       {\mbox{$p$}\xspace}
\def\pt         {\mbox{$p_{\rm T}$}\xspace}

\def\geant      {\mbox{\textsc{Geant4}}\xspace}

\def\pythia     {\mbox{\textsc{Pythia}}\xspace}

\def\tell1  {TELL1\xspace}
\def\ukl1   {UKL1\xspace}

\newcommand{\ie}{\mbox{\itshape i.e.}\xspace}

\usepackage{cite} 
\usepackage{mciteplus}

\usepackage{longtable} 

\begin{document}

\renewcommand{\thefootnote}{\fnsymbol{footnote}}
\setcounter{footnote}{1}


\begin{titlepage}
\pagenumbering{roman}

\vspace*{-1.5cm}
\centerline{\large EUROPEAN ORGANIZATION FOR NUCLEAR RESEARCH (CERN)}
\vspace*{1.5cm}
\noindent
\begin{tabular*}{\linewidth}{lc@{\extracolsep{\fill}}r@{\extracolsep{0pt}}}
\ifthenelse{\boolean{pdflatex}}
{\vspace*{-2.7cm}\mbox{\!\!\!\includegraphics[width=.14\textwidth]{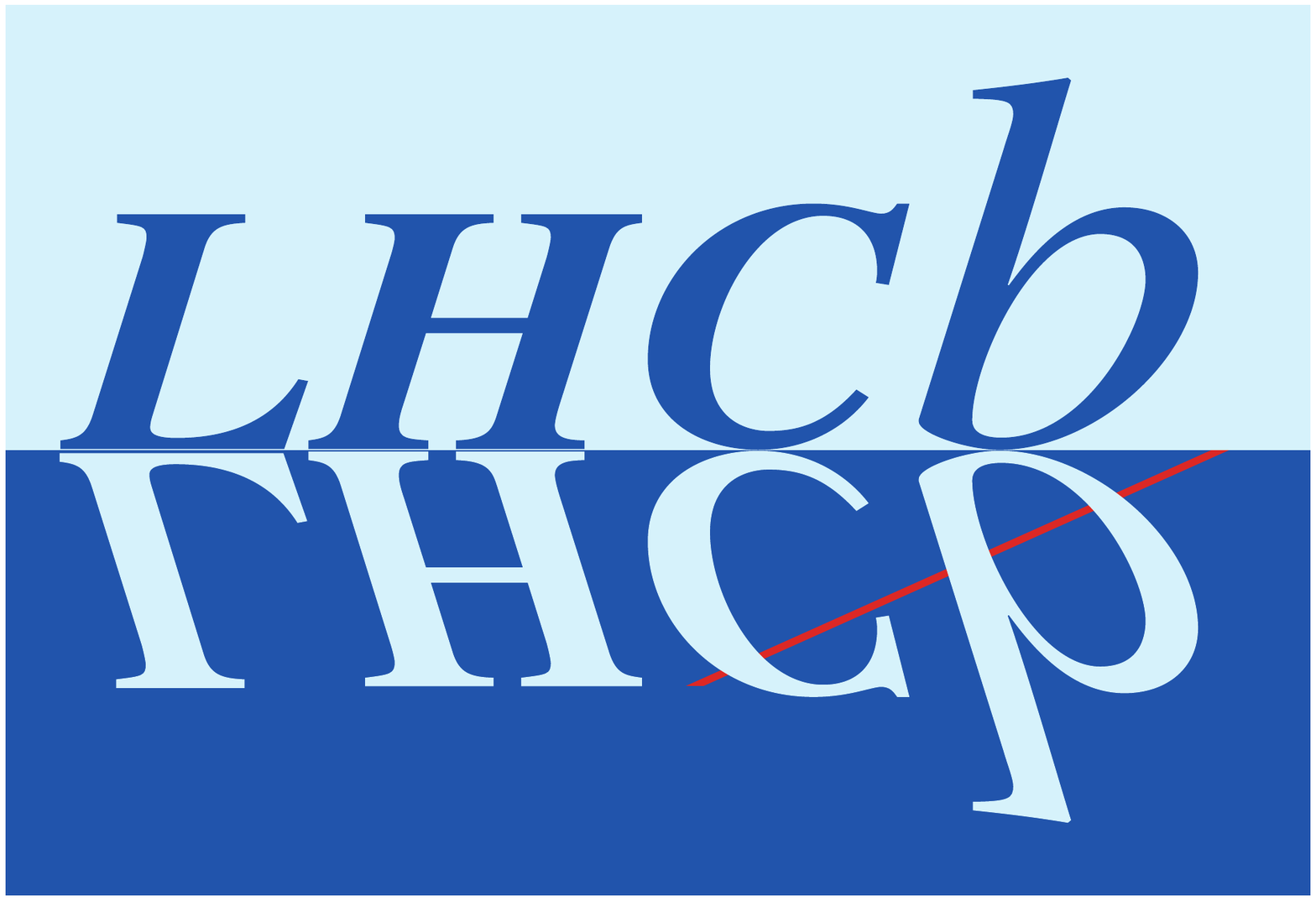}} & &}%
{\vspace*{-1.2cm}\mbox{\!\!\!\includegraphics[width=.12\textwidth]{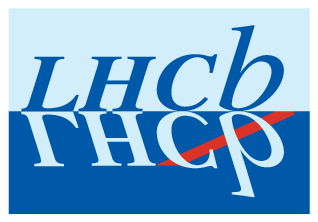}} & &}%
\\
 & & CERN-EP-2016-188 \\  
 & & LHCb-PAPER-2016-014 \\  
 & & September 8, 2016 \\ 
 & & \\
\end{tabular*}

\vspace*{2.0cm}

{\normalfont\bfseries\boldmath\huge
  \begin{center}
        Search for Higgs-like bosons decaying into long-lived exotic particles\\
\end{center}
}

\vspace*{2.0cm}

\begin{center}
The LHCb collaboration\footnote{Authors are listed at the end of this paper.}
\end{center}

\vspace{\fill}

\begin{abstract}
  \noindent
  A search is presented for massive long-lived particles,
  in the 20--60~\gevcc mass range with lifetimes between 5 and 100\ps.
  The dataset used corresponds to 0.62\invfb of proton-proton collision data
  collected by the LHCb detector at $\sqs=7\tev$.
  The particles are assumed to be pair-produced by the decay of
  a Higgs-like boson with mass between 80 and 140\gevcc.
  No excess above the background expectation is observed and limits
  are set on the production cross-section as
  a function of the long-lived particle mass and lifetime and of the Higgs-like boson mass.
\end{abstract}

\vspace*{2.0cm}

\begin{center}
  Published on Eur.~Phys.~J.~C 
\end{center}

\vspace{\fill}

{\footnotesize 
\centerline{\copyright~CERN on behalf of the \lhcb collaboration, licence \href{http://creativecommons.org/licenses/by/4.0/}{CC-BY-4.0}.}}
\vspace*{2mm}

\end{titlepage}


\newpage
\setcounter{page}{2}
\mbox{~}
%
%
%
%

\cleardoublepage


\renewcommand{\thefootnote}{\arabic{footnote}}
\setcounter{footnote}{0}




\pagestyle{plain} 
\setcounter{page}{1}
\pagenumbering{arabic}


%

\newcommand{\mycomment}[1]{}
\newcommand{\sg}{\tilde{\mathrm{g}}}
\newcommand{\sG}{\tilde{\mathrm G}}
\newcommand{\chino}{\tilde{\chi}}
\newcommand{\chio}{\chino^{0}}
\newcommand{\khi}{\ensuremath{\chio_{1}}\xspace}
\newcommand{\chioi}{\chio_{i}}
\newcommand{\chipm}{\chino^{\pm}}
\newcommand{\chip}{\chino^{+}}
\newcommand{\chim}{\chino^{-}}
\newcommand{\chipmi}{\chino^\pm_i}          

\newcommand{\h}{\mathrm{h}}
\newcommand{\Ho}{\H^0}
\newcommand{\ho}{\h^0}

\newcommand{\LLP}{\rm{LLP}}
\newcommand{\LLPs}{\rm{LLPs}}

\newcommand{\RXY}{\ensuremath{R_{\rm xy}}\xspace}

\newcommand{\bone}{\ensuremath{\rm Bkg_1}\xspace}
\newcommand{\btwo}{\ensuremath{\rm Bkg_2}\xspace}
\newcommand{\sfour}{\ensuremath{\rm Sel_2}\xspace}
\newcommand{\ssix}{\ensuremath{\rm Sel_1}\xspace}

\newcommand{\ntrak}{\ensuremath{N_{\rm track}}\xspace}
\newcommand{\mllp}{\ensuremath{m_{\LLP}}\xspace}
\newcommand{\taullp}{\ensuremath{\tau_{\LLP}}\xspace}
\newcommand{\mhzero}{\ensuremath{m_{\rm h^0}}\xspace}
\newcommand{\sigr}{\ensuremath{\sigma_{\rm R}}\xspace}
\newcommand{\sigz}{\ensuremath{\sigma_{\rm Z}}\xspace}

\newcommand{\ntrakmin}{\ensuremath{N^{\rm track}_{\rm min}}\xspace}
\newcommand{\mllpmin}{\ensuremath{m^{\rm LLP}_{\rm min}}\xspace}
\newcommand{\sigrmax}{\ensuremath{\sigma^{\rm R}_{\rm max}}\xspace}
\newcommand{\sigzmax}{\ensuremath{\sigma^{\rm Z}_{\rm max}}\xspace}

\newcommand\crule[3][black]{\textcolor{#1}{\rule{#2}{#3}}}

\newcommand{\plhcb}{{\makebox[1.1\width]{LHCb}}}
\newcommand{\plhcbs}{ \scalebox{0.9}{{\makebox[1.1\width]{LHCb}}} }

\newcommand{\kbb}{\khi\rightarrow\nu\b\bbar}
\newcommand{\kmm}{\khi\rightarrow\nu\mu^+\mu^-}
\newcommand{\kmq}{\khi\rightarrow\mu^\pm\q\q'}
\newcommand{\higgskhi}{\ho\rightarrow\khi\khi}
\newcommand{\higgskhijets}{\ho\rightarrow\khi\khi\rightarrow 6\ \mbox{jets}}
\newcommand{\khijets}{\khi\rightarrow 3\ \mbox{jets}}
\newcommand{\Zkhijets}{\Zo\rightarrow\khi\khi\rightarrow 6\ \mbox{jets}}
\newcommand{\Zkhi}{\Zo\rightarrow\khi\khi}
\newcommand{\hssbbbb}{\ho\rightarrow\s\s\rightarrow\b\bbar\b\bbar}

\def\LSP{LSP}                             
\def\tanb{$\tan{\beta}$}                  
\def\mzero{$\m_0$}                        
\def\mhalf{$\m_{1/2}$}                     
\def\mazero{$\m_{\rm A_0}$}                    
\def\sgnmu{sign($\mu$)}                   

\newcommand{\LL}{\ensuremath{0.62}\,}
\newcommand{\LLE}{\ensuremath{1.7}\,}
\newcommand{\SYST}{\ensuremath{20.5}\,}

\section{Introduction}

The standard model of particle physics (SM) has shown great success in describing physics processes at       
very short distances.  Nevertheless, open questions remain, such as the hierarchy  problem,      
the imprecise unification of gauge couplings, and the absence of candidates for dark matter.         
Considerable efforts have been made to address these issues, resulting in a large variety of        
models. 
Supersymmetry (SUSY), in which the strong and electroweak forces are           
unified at a renormalisation scale near the Planck scale, provides a possible solution for the      
hierarchy problem;
the minimal supersymmetric standard model (MSSM) is the simplest,
phenomenologically viable realisation of SUSY~\cite{MSSM,MSSMM}.

The present study focuses on a subset of models featuring massive
long-lived particles (LLP) with a measurable flight distance.
We concentrate on scenarios in which the LLP decays hadronically
in the LHCb vertex detector, travelling distances which can be larger 
than those of typical \bquark hadrons.

A large number of LLP searches have been performed by the experiments at the LHC and Tevatron,
mainly using the Hidden Valley framework~\cite{hv2}
as a benchmark model~\cite{D0-HV, CDF-HV, CMS-2015, ATLAS-HV, ATLAS-neutralino-lepton-2012}.
Hidden Valley processes have also been sought by LHCb~\cite{LHCb-PAPER-2014-062},
which is able to explore the forward rapidity region
only partially covered by other LHC experiments.
In addition, it is able to trigger on particles with low transverse momenta,
allowing the experiment to probe relatively small LLP masses.

The event topology considered in this study is quite different from that of Hidden Valley models.
The  minimal supergravity model (mSUGRA) realisation of the MSSM is used as a benchmark model
with baryon number violation~\cite{redfintun},
as suggested in  Refs.~\cite{KaplanLHCb,KaplanDisplaced2012}.
Here a Higgs-like boson produced in $pp$ collisions decays into two LLPs (neutralinos),
subsequently decaying into three quarks each.
The Higgs-like particle mass ranges from 80\gevcc up to 140\gevcc,
covering the mass of the scalar boson discovered by the ATLAS and CMS experiments~\cite{Aad:2012tfa,Chatrchyan:2012ufa}.
The explored LLP lifetime range of 5--100\ps is higher than the typical \bquark hadron lifetime,
and corresponds to an average flight distance of up to  30\cm,
which is inside the LHCb vertex detector region.
The LLP mass range considered is between 20\gevcc and 60\gevcc.

\section{Detector description}
\label{sec:Detector}
The \lhcb detector~\cite{Alves:2008zz,LHCb-DP-2014-002} is a single-arm forward
spectrometer covering the \mbox{pseudorapidity} range $2<\eta <5$,
designed for the study of particles containing \bquark or \cquark
quarks.
The detector includes a high-precision tracking system
consisting of a silicon-strip vertex detector surrounding the $pp$
interaction region (\velo),    
a large-area silicon-strip detector located
upstream of a dipole magnet with a bending power of about
$4{\mathrm{\,Tm}}$, and three stations of silicon-strip detectors and straw
drift tubes, 
placed downstream of the magnet.
The tracking system provides a measurement of the momentum, \ptot, of charged particles with
a relative uncertainty that varies from 0.5\% at low momentum to 1.0\% at 200\gevc.
The minimum distance of a track to a primary vertex (PV), the impact parameter,
is measured with a resolution of $(15+29/\pt)\mum$,
where \pt is the component of the momentum transverse to the beam, in\,\gevc.
Different types of charged hadrons are distinguished using information
from two ring-imaging Cherenkov detectors. 
Photons, electrons and hadrons are identified by a calorimeter system consisting of
scintillating-pad and preshower detectors, an electromagnetic
calorimeter and a hadronic calorimeter. Muons are identified by a
system composed of alternating layers of iron and multiwire
proportional chambers.
The online event selection is performed by a trigger~\cite{LHCb-DP-2012-004}, 
which consists of a hardware stage, L0, based on information from the calorimeter and muon
systems, followed by two software stages,  HLT1 and HLT2, which run a simplified version
of the offline event reconstruction.

\section{Event generation and detector simulation}\label{sec:evtgen}

Various simulated event samples are used in this analysis.
The $pp$ collisions are generated with \pythia~6~\cite{Sjostrand:2006za}. 
The process simulated is $\rm \ho \rightarrow \khi \khi$, where the
Higgs-like boson of mass \mhzero is produced via gluon-gluon fusion,
with the parton density function taken from CTEQ6L~\cite{cteq6l}.
The neutralino \khi is an LLP of mass \mllp and lifetime $\tau_{\LLP}$, which decays into three quarks via the mSUGRA
baryon number violating process available in \pythia.
The corresponding decay flavour structure
for  the neutralino with a mass of $48\gevcc$ is 18.5\% for each of the
combinations with a \bquark quark ($udb$, $usb$, $cdb$, $csb$), and 13\% for each $udq$ and $cdq$,
where $q$ is not a \bquark quark, \ie about 75\% of LLPs have a \bquark quark in the decay.
This fraction becomes 70\% for $\mllp=20\gevcc$.

Two separate detector simulations are used,
a full simulation where the interaction of the generated particles with the detector
is based on \geant~\cite{Allison:2006ve, *Agostinelli:2002hh},
and a fast simulation.
In \geant, the detector and its response are implemented as described in Ref.~\cite{LHCb-PROC-2011-006}.
Signal models for a representative set of theoretical parameters
have been generated and fully simulated (Appendix~\ref{app:gen}, Table~\ref{tab:models-full}).
In the remainder of this paper, the following nomenclature is chosen: a
prefix ``BV'', indicating baryon number violation, is followed by the LLP mass in \gevcc and
lifetime, and the prefix ``mH'' followed by the \mhzero value in \gevcc.
Most of the fully simulated models have  $\mhzero$=114\gevcc, which is in the middle of
the chosen Higgs-like particle mass range.
Only events with at least one \khi
in the pseudorapidity region $1.8<\eta<5.0$ are processed by \geant,
corresponding to about 30\% of the generated events.

The fast simulation is used to cover a broader parameter space 
of the theoretical models.
Here the charged particles from the $\rm \ho \rightarrow \khi \khi$ process falling in the
geometrical acceptance of the detector are processed by the vertex reconstruction
algorithm.
The fast simulation is validated by comparison with the full simulation.
The detection efficiencies predicted by the full
and the fast simulation differ by less than 5\%
for all the signal models.
The distributions for mass, momentum and transverse momentum of the reconstructed LLP,
and for the reconstructed vertex position coincide.

Events with direct production of charm, bottom and top quarks
are considered as sources of background. Samples of such events were produced
and fully simulated.
In particular, $17 \times 10^6$ inclusive \bbbar events ($9 \times 10^6$ inclusive \ccbar events) were produced
with at least two \bquark hadrons (\cquark hadrons) in 
$1.5<\eta<5.0$, and half a million \ttbar events with at least one muon in the acceptance.

\section{Event selection and signal determination}
This analysis searches for events with pairs of displaced high-multiplicity vertices.
The main background is due to secondary interactions
of particles with the detector material.
These events are discarded by a material veto, which rejects vertices in regions
occupied by detector material~\cite{LHCb-PAPER-2012-023}.
The remaining candidates are found to be compatible with \bbbar events.

From simulation, LLP candidates within the detector acceptance are selected by the L0 and HLT1 triggers
with an efficiency of more than  85\%.
The simulation indicates that the trigger activity is dominated by
the hadronic component of the signal expected from high multiplicity events.
In HLT2, primary vertices and displaced vertices are reconstructed from
charged tracks~\cite{LHCb-PUB-2014-044}.
Genuine PVs are identified by a small radial distance from the beam axis, $\RXY<0.3$~mm,
and must have at least 10 tracks, including at least one forward track
(\ie in the direction of the spectrometer) and one backward track.
Once the set of PVs is identified, all other reconstructed vertices
are candidates for the decay position of LLPs.
The preselection requires at least one PV in the event and two LLP candidates.
The LLP candidates must have at least four forward tracks, no backward tracks, and 
a minimum invariant mass reconstructed from charged tracks larger than 3.5\gevcc
for one candidate, and larger than 4.5\gevcc for the other.
In addition, the two secondary vertices must have $\RXY > 0.4\mm $ and pass the material veto. 

The preselection criteria drastically suppress the hadronic background.
Only 37 events (74 LLP candidates) survive from
the simulated set of $17.1 \times 10^6$ \bbbar events generated in the LHCb acceptance, corresponding
to an integrated luminosity of 0.3\invpb.
Three simulated \ccbar events pass the selection.  They contain \bquark hadrons and hence
belong to the category of inclusive \bbbar, which is also the case of the two
surviving \ttbar events.
From the 0.62\invfb data sample, $42.9 \times 10^3$ events are selected.
The \bbbar cross-section value measured by LHCb,
$ 288 \pm 4 \pm 48$ $\mu$b~\cite{ LHCb-PAPER-2011-003, LHCb-PAPER-2010-002},
predicts  $(76 \pm 22)\times10^3$ events,
$1.8\pm0.5$ times  the yield observed in data.
The estimate uses the next-to-leading-order POWHEG calculation~\cite{POWHEG} to correct \pythia,
and the detection efficiency obtained from the simulated events.
The measured yield has also been compared to the rate observed in LHCb by
a dedicated inclusive \bbbar analysis, based on a topological trigger~\cite{LHCb-PUB-2011-002}.
The consistency with the \bbbar background is verified within a statistical precision of 10\%.

\begin{figure}[h!]
\begin{center}
\includegraphics[width=0.5\linewidth]{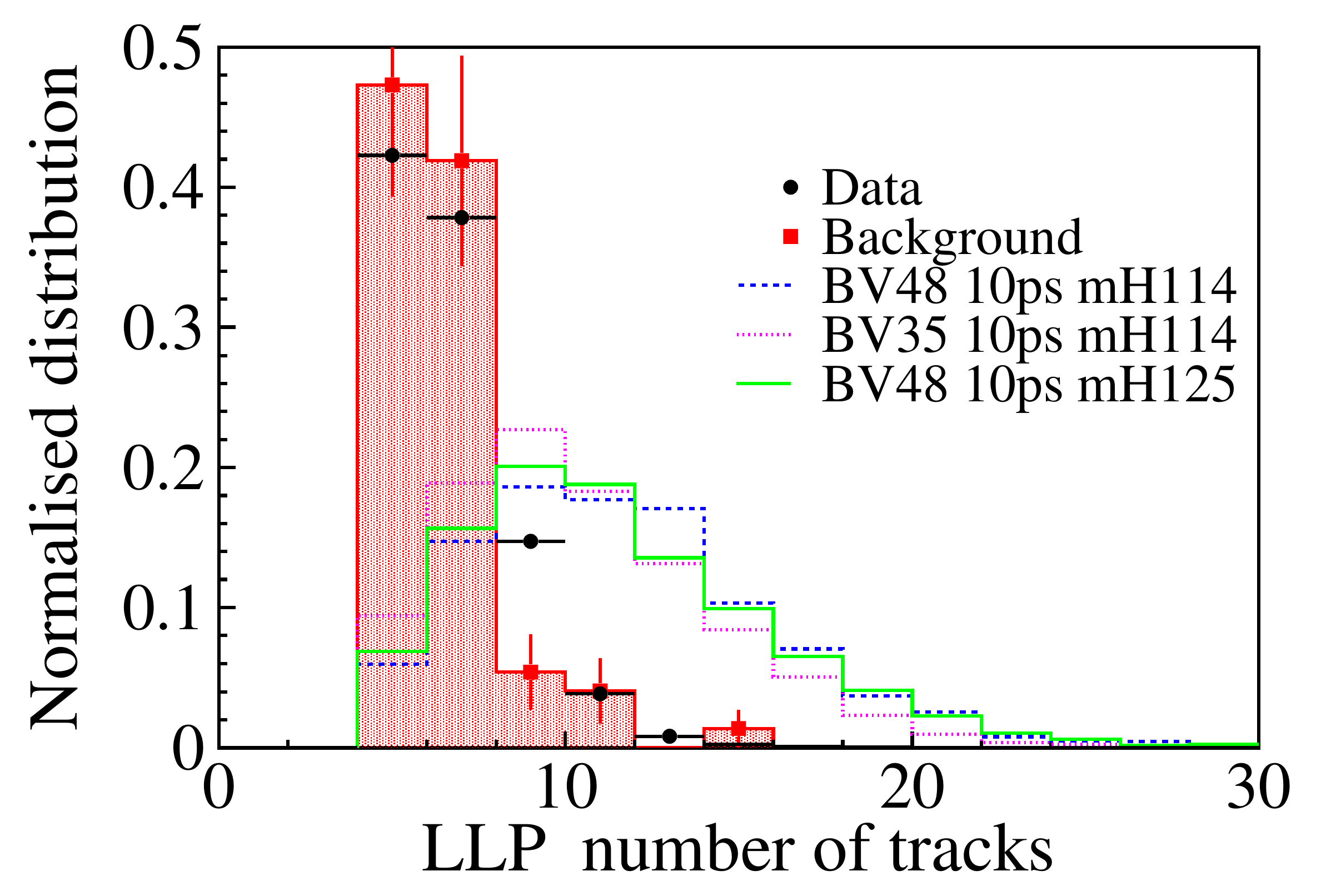}\put(-94,134){(a)}\put(-47,132){\plhcb}
\includegraphics[width=0.5\linewidth]{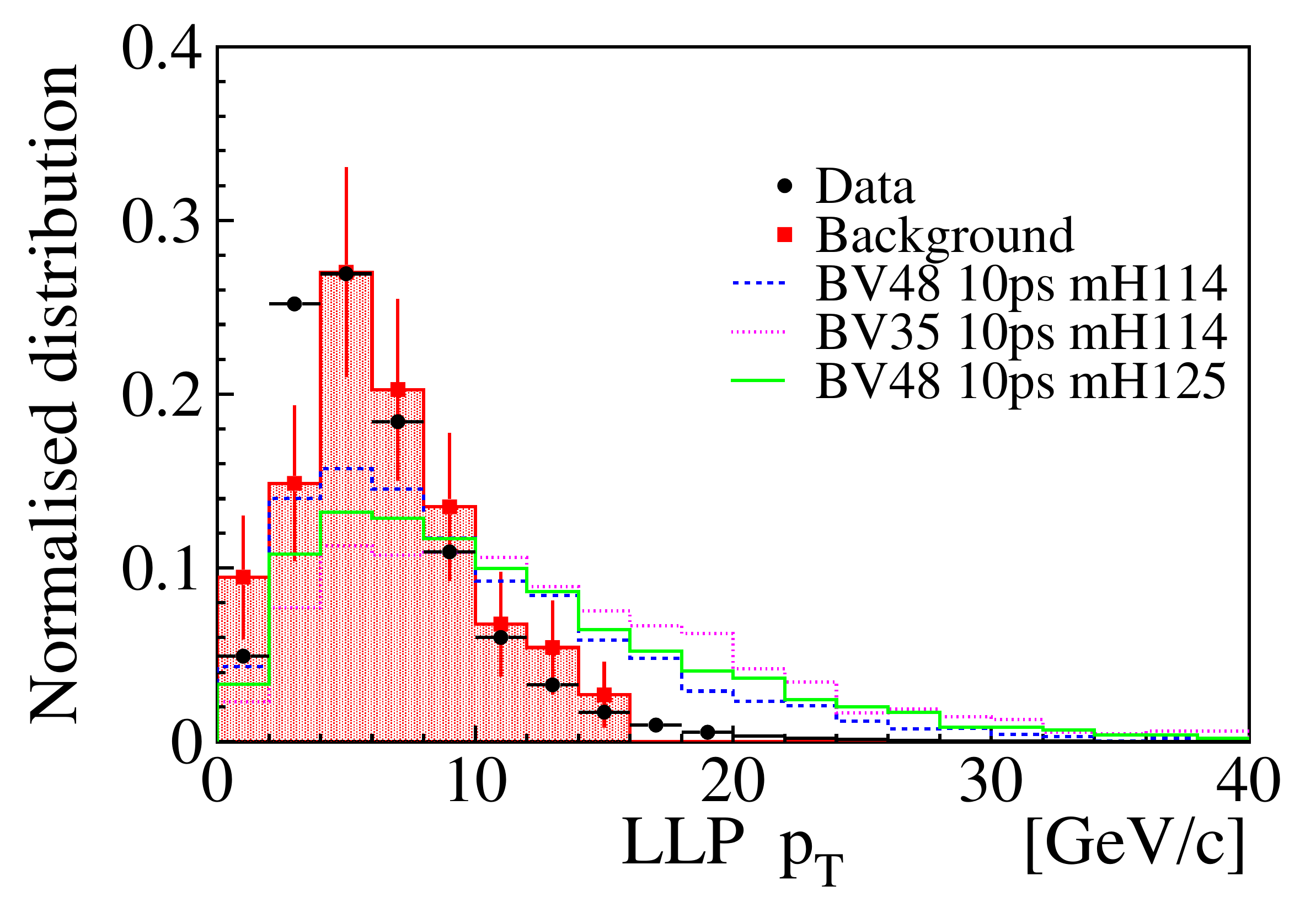}\put(-94,136){(b)}\put(-47,132){\plhcb}\\
\includegraphics[width=0.5\linewidth]{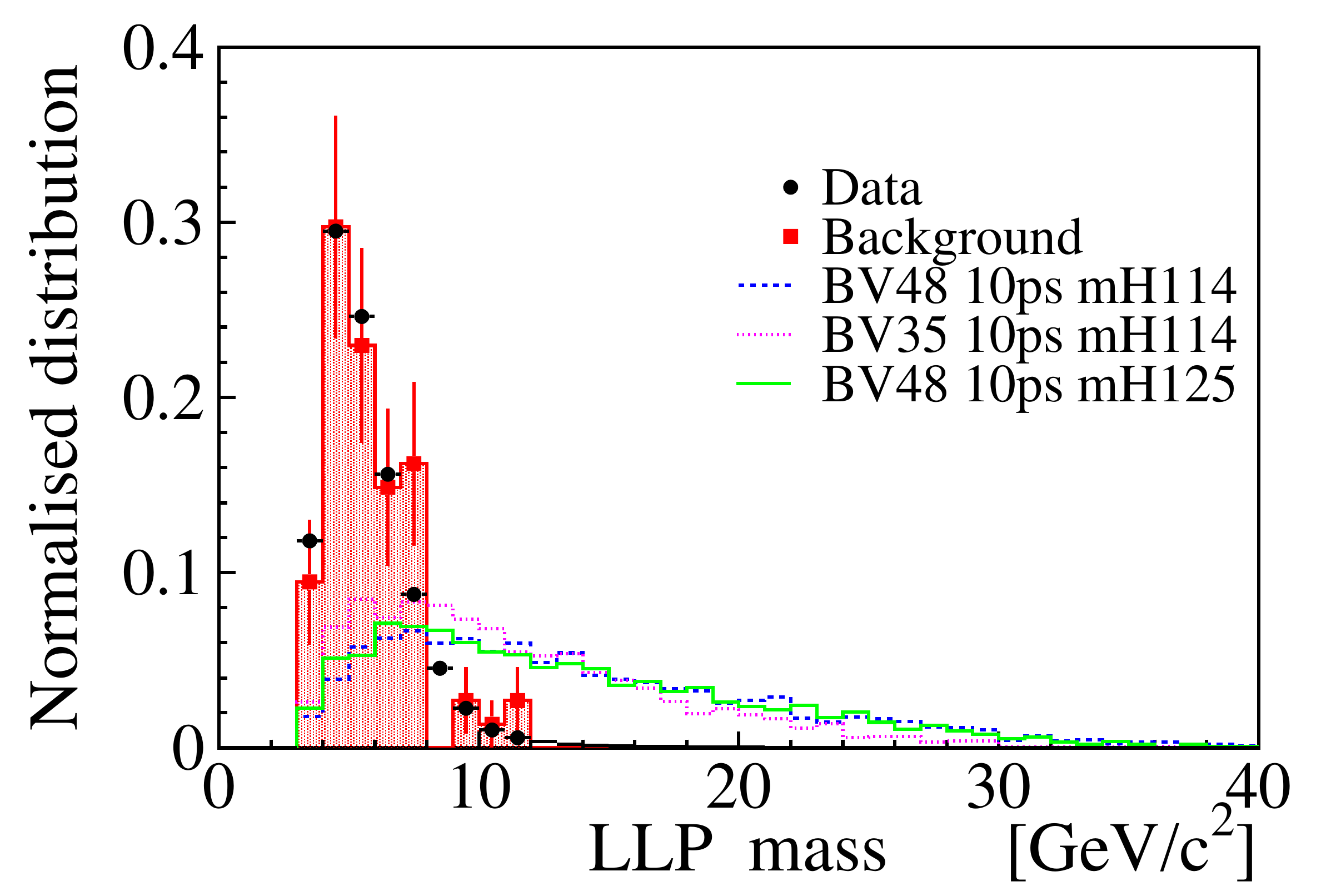}\put(-94,134){(c)}\put(-47,132){\plhcb}
\includegraphics[width=0.5\linewidth]{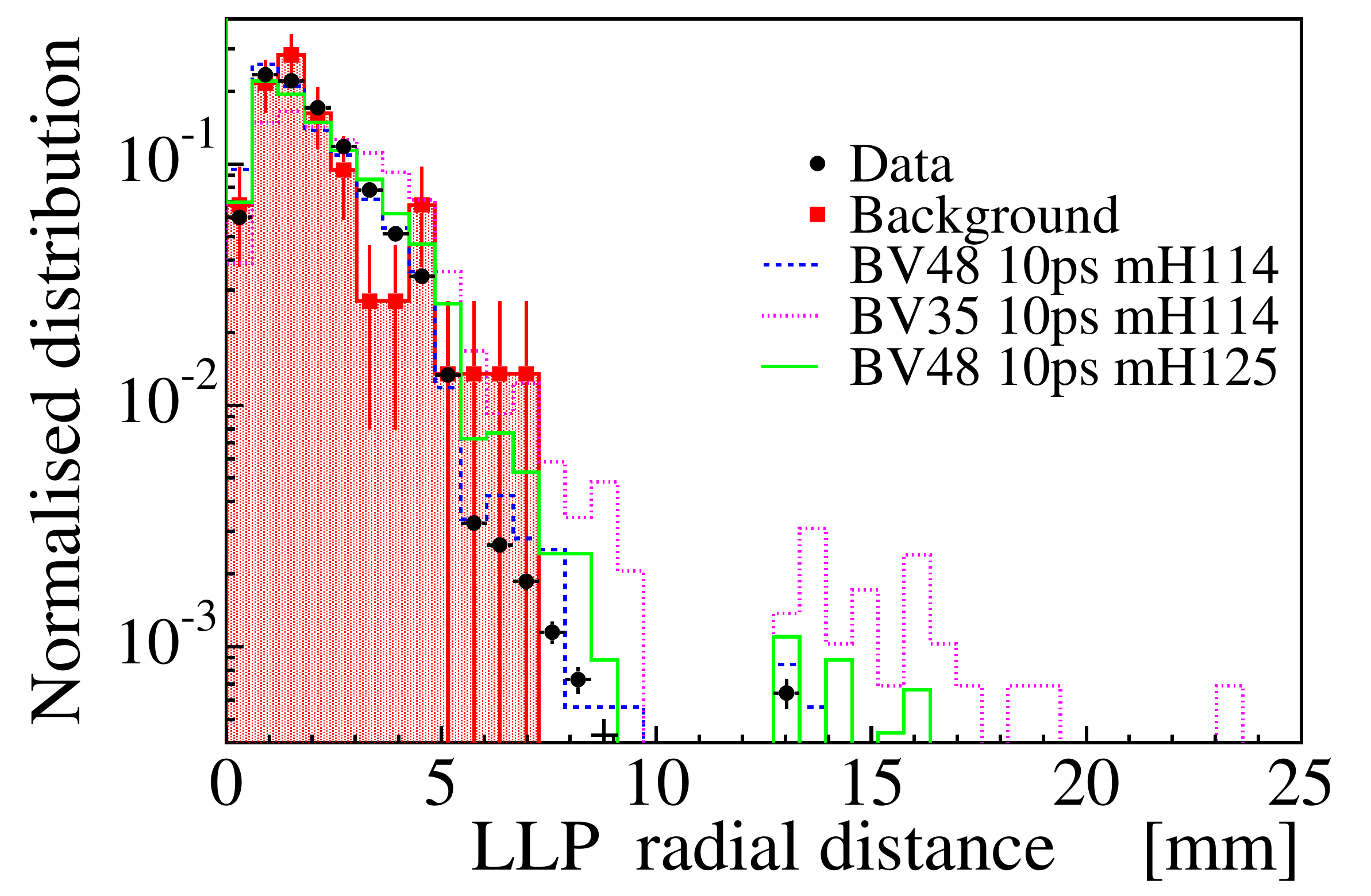}\put(-94,134){(d)}\put(-47,132){\plhcb}\\
\includegraphics[width=0.505\linewidth]{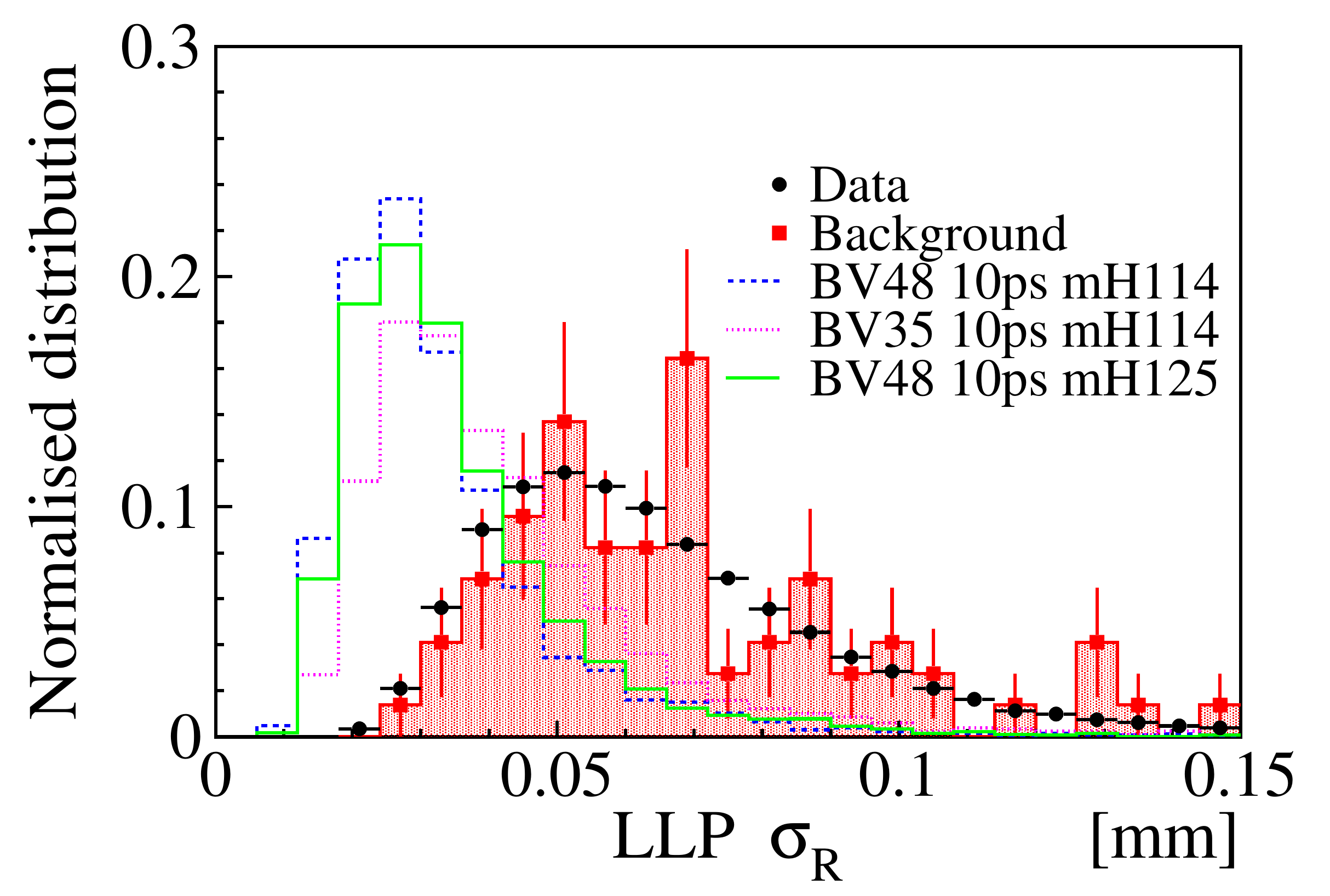}\put(-97,134){(e)}\put(-50,132){\plhcb}
\includegraphics[width=0.49\linewidth]{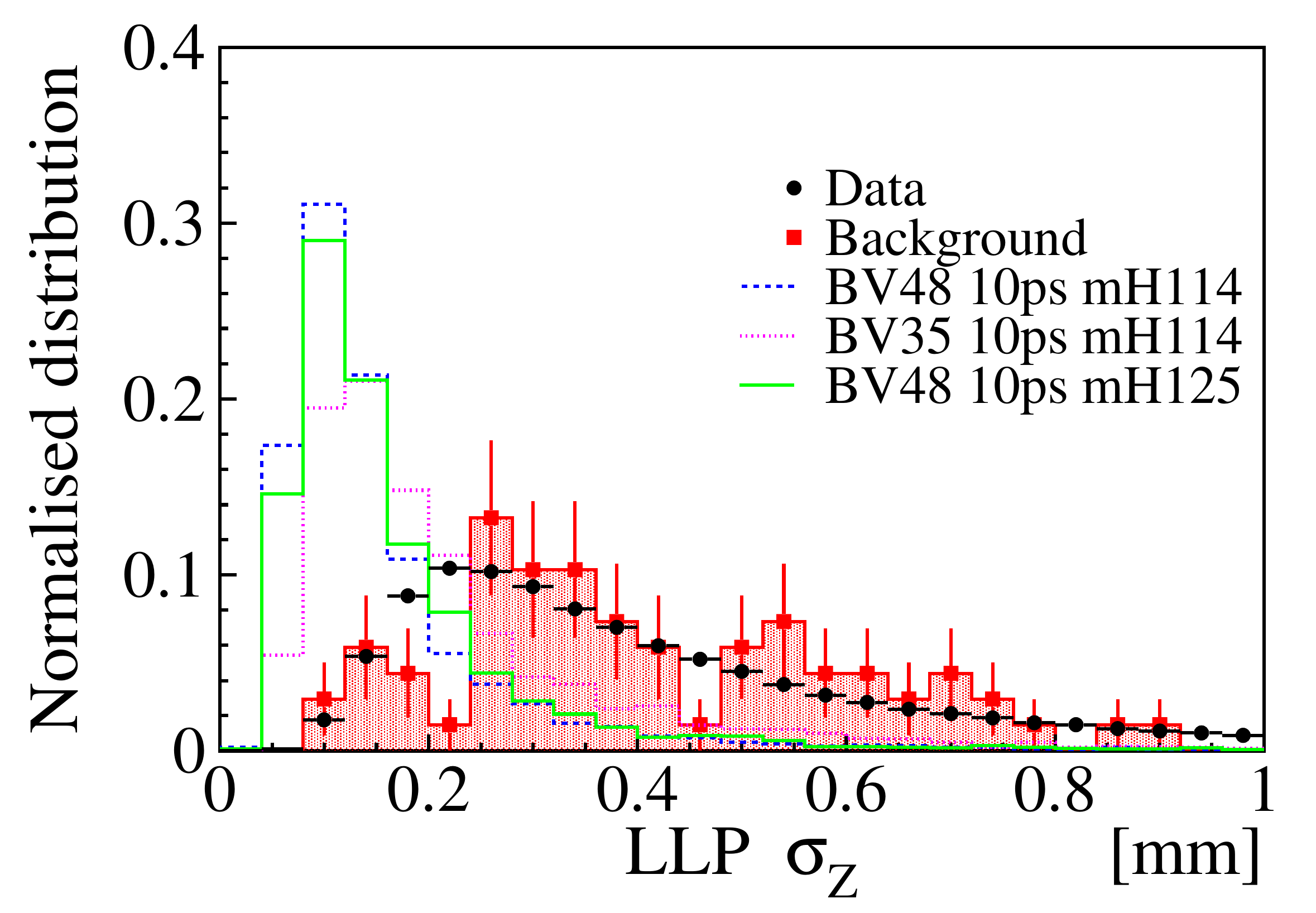}\put(-94,134){(f)}\put(-45,131){\plhcb}\\
\caption{\small
 Data (black dots) and simulated distributions after preselection normalised to unit integral.
 There are two LLP candidates per event. The simulated \bbbar background is shown by the filled
 red histograms with error bars. The dashed (blue), dotted (purple) and solid (green) lines
 are distributions for fully simulated signal models.  The subplots show (a) number of tracks
 used to reconstruct the LLP candidates, (b) LLP transverse momentum, (c) LLP invariant
 mass, (d) radial distance, \RXY, (e) uncertainty of the radial position, \sigr, and (f)
 uncertainty of the longitudinal position, \sigz, of the LLP vertex. }
\label{fig:LLP}
\end{center}
\end{figure}

\begin{figure}
\begin{center}
\includegraphics[width=0.5\linewidth]{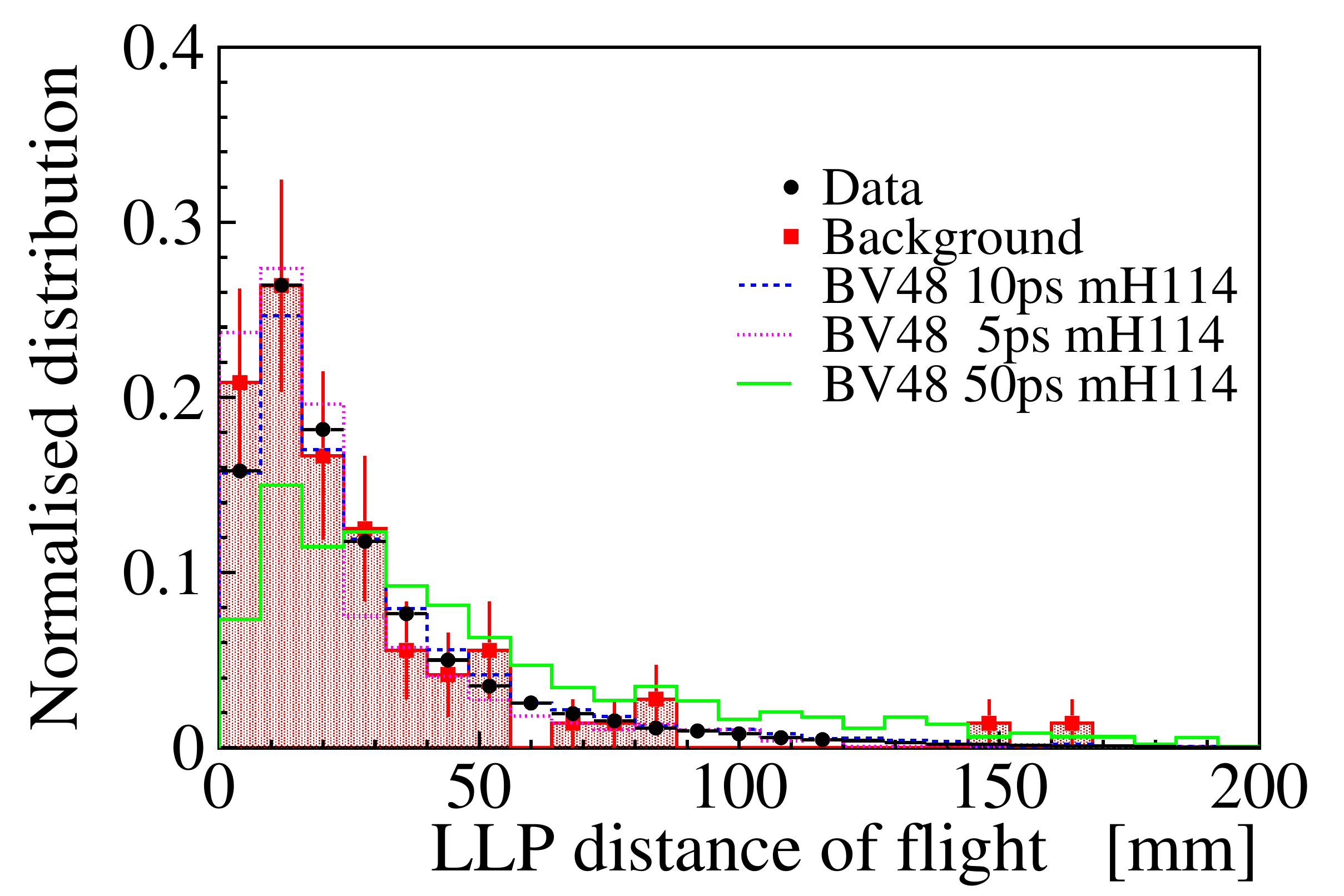}\put(-94,133){(a)}\put(-48,131){\plhcb}
\includegraphics[width=0.5\linewidth]{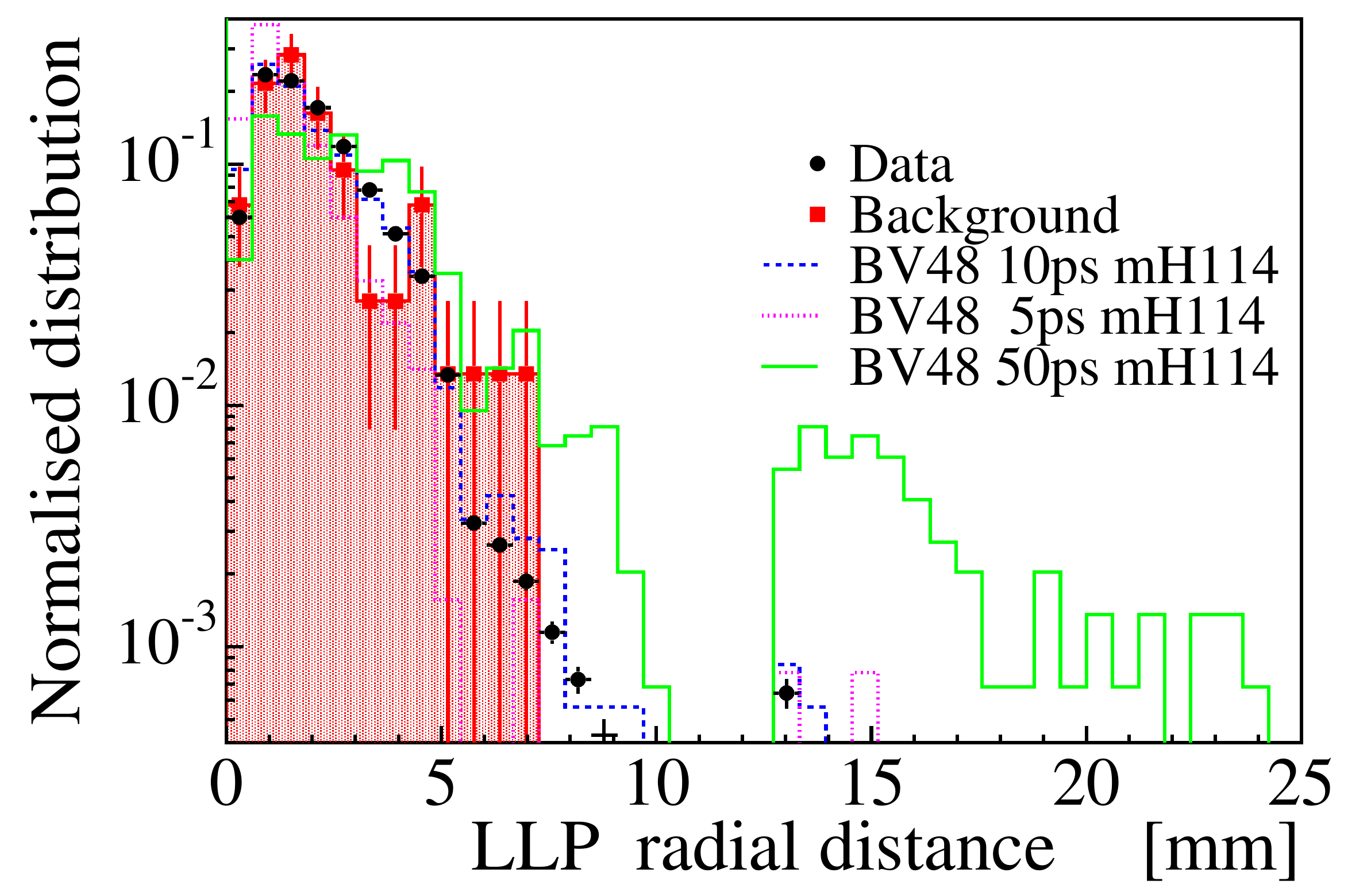}\put(-94,134){(b)}\put(-46,132){\plhcb}\\
\caption{\small
    Distributions for
  (a) the LLP distance of flight from the PV, and, (b), the radial distance of the LLP vertex, \RXY.
    The fully simulated signal models are chosen with LLP lifetimes of 5, 10, and 50\ps.
    Symbols are defined as in  Fig.~\ref{fig:LLP}.
  }
\label{fig:LLP-t}
\end{center}
\end{figure}

The shapes of the distributions of the relevant observables are compatible
with the \bbbar background.
Figure~\ref{fig:LLP} compares the distributions for the LLP candidates
taken from data and from simulated \bbbar events.
The distributions for three fully simulated signal models are also shown.
The mass and the \pt values are calculated assuming the pion mass for each charged track.
Figure~\ref{fig:LLP}(d)  presents the radial distribution of the displaced vertices;
the drop in the number of candidates  with a vertex above $\RXY\sim 5\mm$ is
due to the material veto.
The variables \sigr and \sigz shown in Figs.~\ref{fig:LLP}(e) and (f) are the position
uncertainties provided by the vertex fit in the transverse distance \RXY
and along the $z$ axis, parallel to the beam.
The values of \sigr and \sigz are larger for the candidates from \bbbar background than for the signal
because light boosted particles produce close parallel tracks, with the consequence that the vertex fit
has larger uncertainties than for the  decay of heavier particles producing more diverging tracks.
Figure~\ref{fig:LLP-t} presents the LLP distance of flight and \RXY distributions compared
to three fully simulated signal models, corresponding to \taullp values of 5, 10, and 50\ps.

\begin{figure}
\begin{center}
\includegraphics[width=0.5\linewidth]{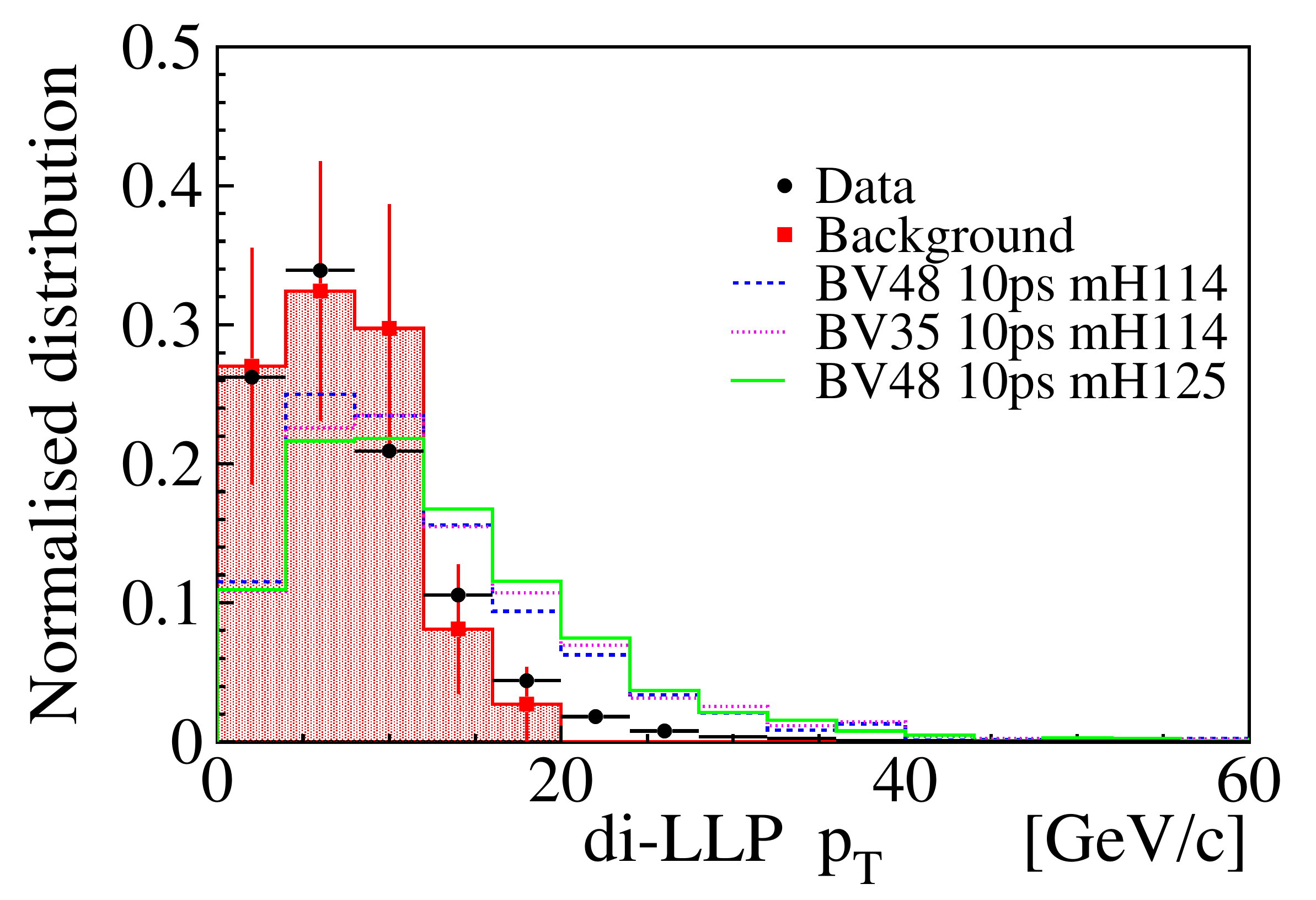}\put(-94,136){(a)}\put(-47,135){\plhcb}
\includegraphics[width=0.5\linewidth]{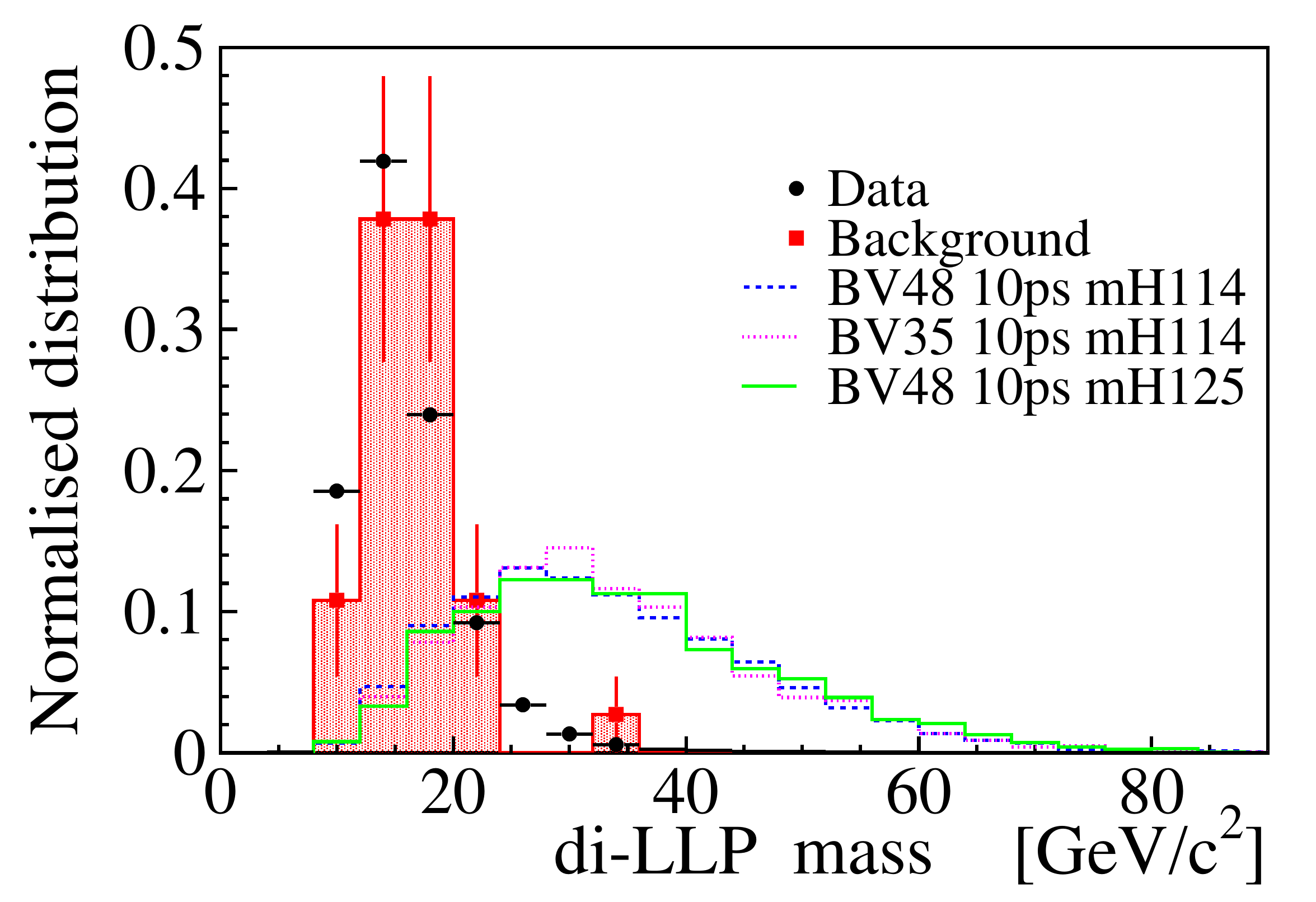}\put(-94,135){(b)}\put(-46,135){\plhcb}\\
\caption{\small
  Distributions for
 (a) the \pt of the Higgs-like candidate, and (b), its invariant mass.
    Symbols are defined as in  Fig.~\ref{fig:LLP}.
}
\label{fig:diLLP}
\end{center}
\end{figure}

The reconstructed four-vectors of the two LLPs in the event are added to form the Higgs-like candidate (di-LLP),
the corresponding invariant mass and \pt distributions are given in  Fig.~\ref{fig:diLLP}.

Further cuts are applied to the preselected data, to increase the statistical sensitivity.
The figure of merit used is 
given by $\epsilon / \sqrt{N_d + 1}$, where $\epsilon$ is
the signal efficiency from simulation for a given selection,
and $N_d$ the corresponding number of candidates found in the data.
The baseline selection (\ssix) is defined by a minimum number of charged tracks on each vertex
$\ntrakmin = 6$,  a minimum reconstructed mass $\mllpmin=6\gevcc$,
and maximum uncertainties from the vertex fit $\sigrmax=0.05 \mm$, and $\sigzmax=0.25\mm$.
All the selections used in this analysis are 
described in Table~\ref{tab:ana-selections}, with the indication of the number of
data events selected for a di-LLP reconstructed mass above 19\gevcc.
Selection \bone is used to model the background in the fit procedure described
in Setion~\ref{sec:signal-extraction}, selections \sfour and \btwo
are used to study systematic effects.

\begin{table}[tb]
\caption{\small
 Definition of the criteria used for the signal determination.
 Selections  \ssix  and \bone are the baseline selections used in the fit,
 \sfour and \btwo are used for the determination of systematic effects.
 The material veto and the requirement $\RXY>0.4$\mm are applied to both LLP candidates.
 The last column gives the number of data events selected, for a di-LLP reconstructed mass above 19\gevcc.
}
\renewcommand{\arraystretch}{1.15}
\begin{center}
\begin{tabular}{cccccr}
  \hline
Selection &  \ntrakmin & \mllpmin & \sigrmax & \sigzmax & $N_{\rm d}$   \\
          &                  & [\gevcc]          & [\mm]  & [\mm]  &     \\
\hline
  \ssix      &   6              & 6    & 0.05   & 0.25         &  157   \\
  \sfour     &   5              & 5    & 0.05   & 0.25         &  387   \\ 
\hline
  \bone      &   4              & 4    & --      & --          &  23.2k    \\ 
  \btwo      &   5              & 5    & --      & --          &  10.1k    \\ 
\hline
\end{tabular}
\label{tab:ana-selections}
\end{center}
\end{table}

\section{Determination of the di-LLP signal}\label{sec:signal-extraction}

The signal yield is determined by a fit of the di-LLP invariant mass,
assuming that the two LLPs are the decay products of a narrow resonance.
This technique is hampered by the difficulty in producing a reliable background model
from simulation, despite the fact that it is reasonable to believe that
only \bbbar events are the surviving SM component.
Therefore, in this analysis the alternative is chosen to infer the background model from data by relaxing
the selection requirements, as given by lines \bone and \btwo of Table~\ref{tab:ana-selections}. 
The comparison of the results obtained with the different signal and background
selections is subsequently used to estimate the systematic effects.

\begin{figure}[tb]
  \begin{center}
  \includegraphics[width=0.64\linewidth]{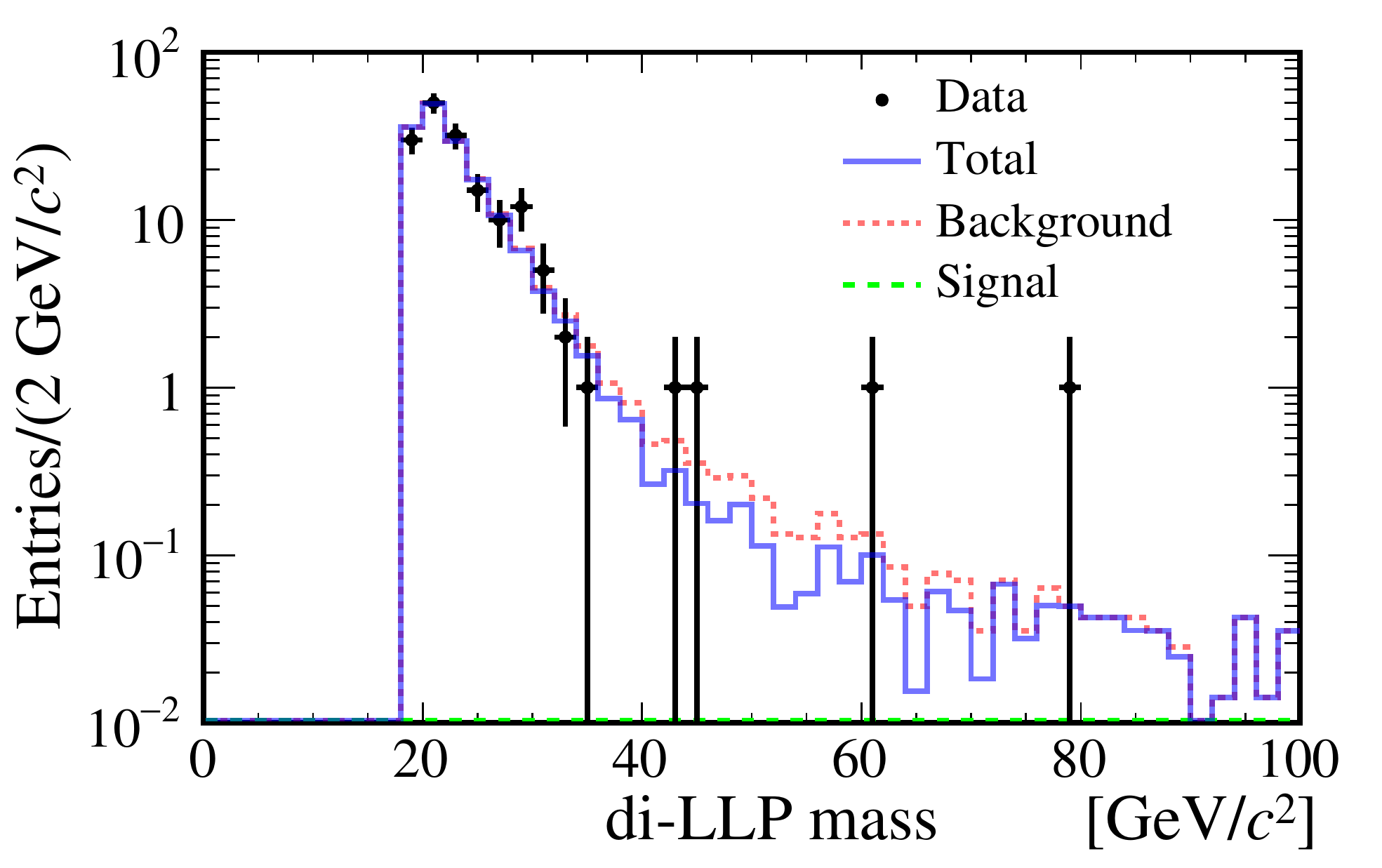}   \put(-160,155){(a)}\put(-58,155){\plhcbs} \\
  \includegraphics[width=0.64\linewidth]{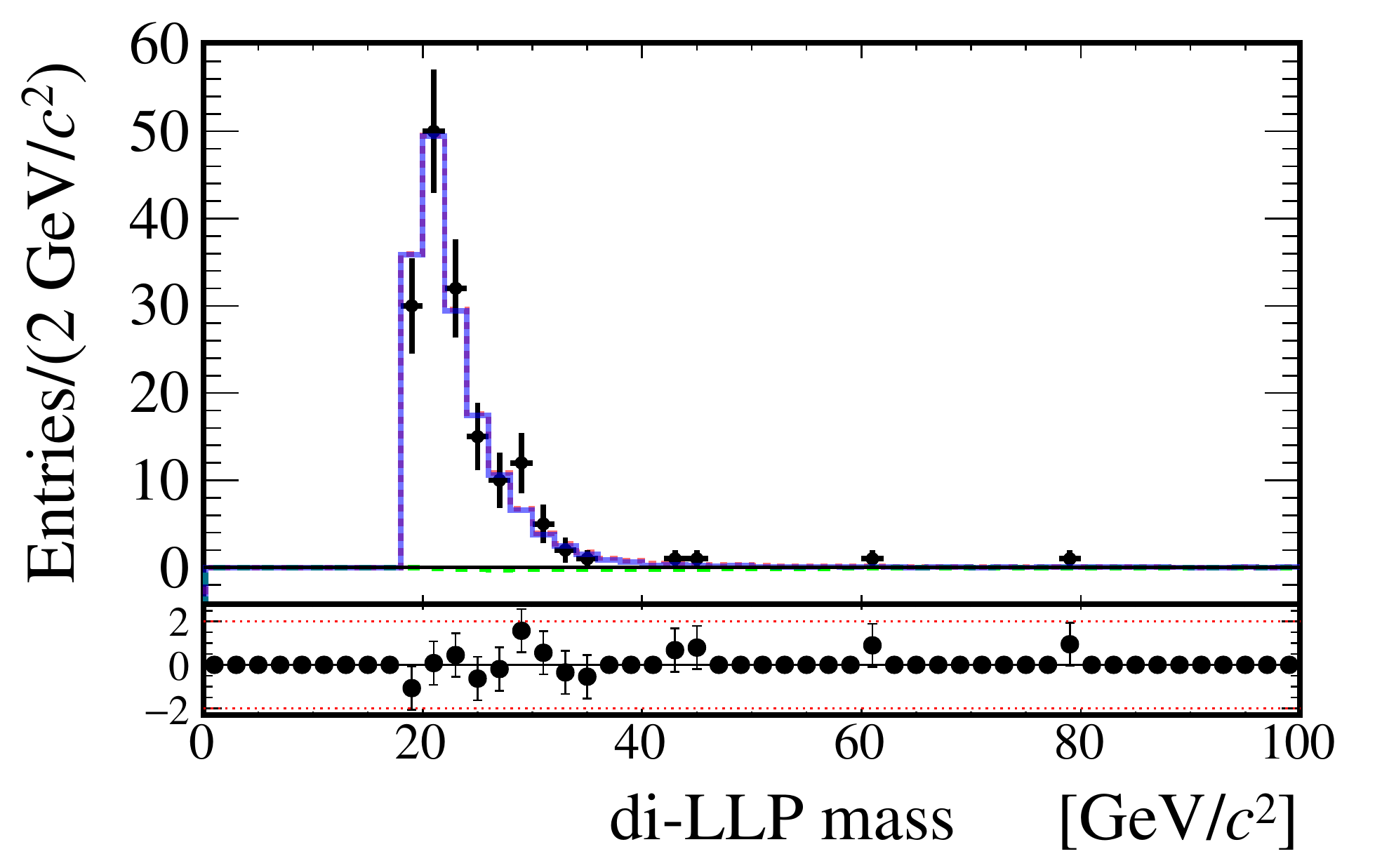} \put(-160,155){(b)}\put(-58,158){\plhcbs} 
  \caption{\small Results of the fit based on the model BV48 10ps mH114.
  In (a) log distribution and (b) linear scale with pull distribution.
 Dots with error bars are the data, the dotted (red) and the dashed (green) histograms
 show the fitted background and signal contributions, respectively.
  The purple histogram is the total fitted distribution. 
}
\label{fig:fit-histopdf-bv48}
\end{center}
\end{figure}

\begin{figure}[tb]
  \begin{center}
  \includegraphics[width=0.64\linewidth]{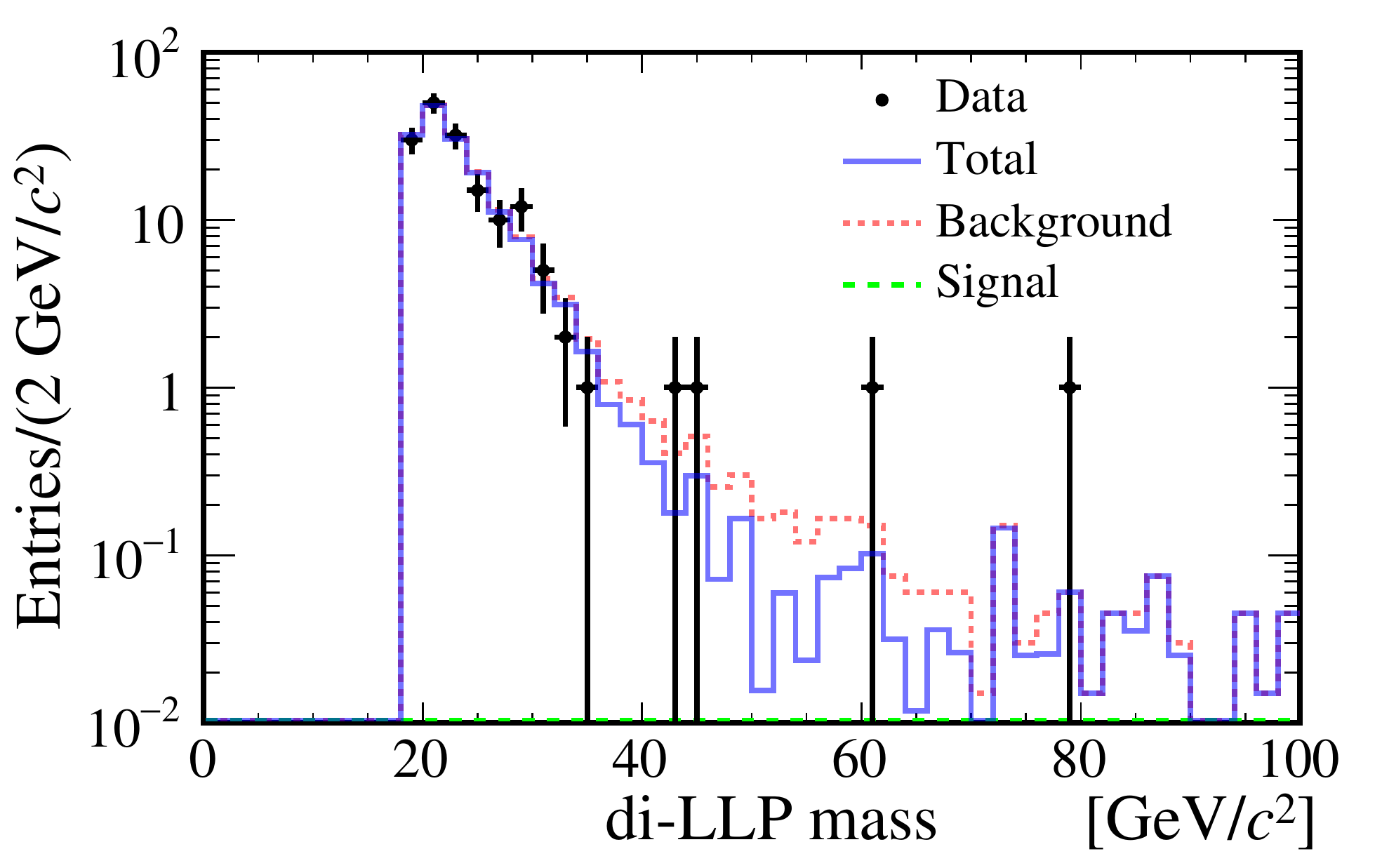}   \put(-160,155){(a)}\put(-58,155){\plhcbs} \\
  \includegraphics[width=0.64\linewidth]{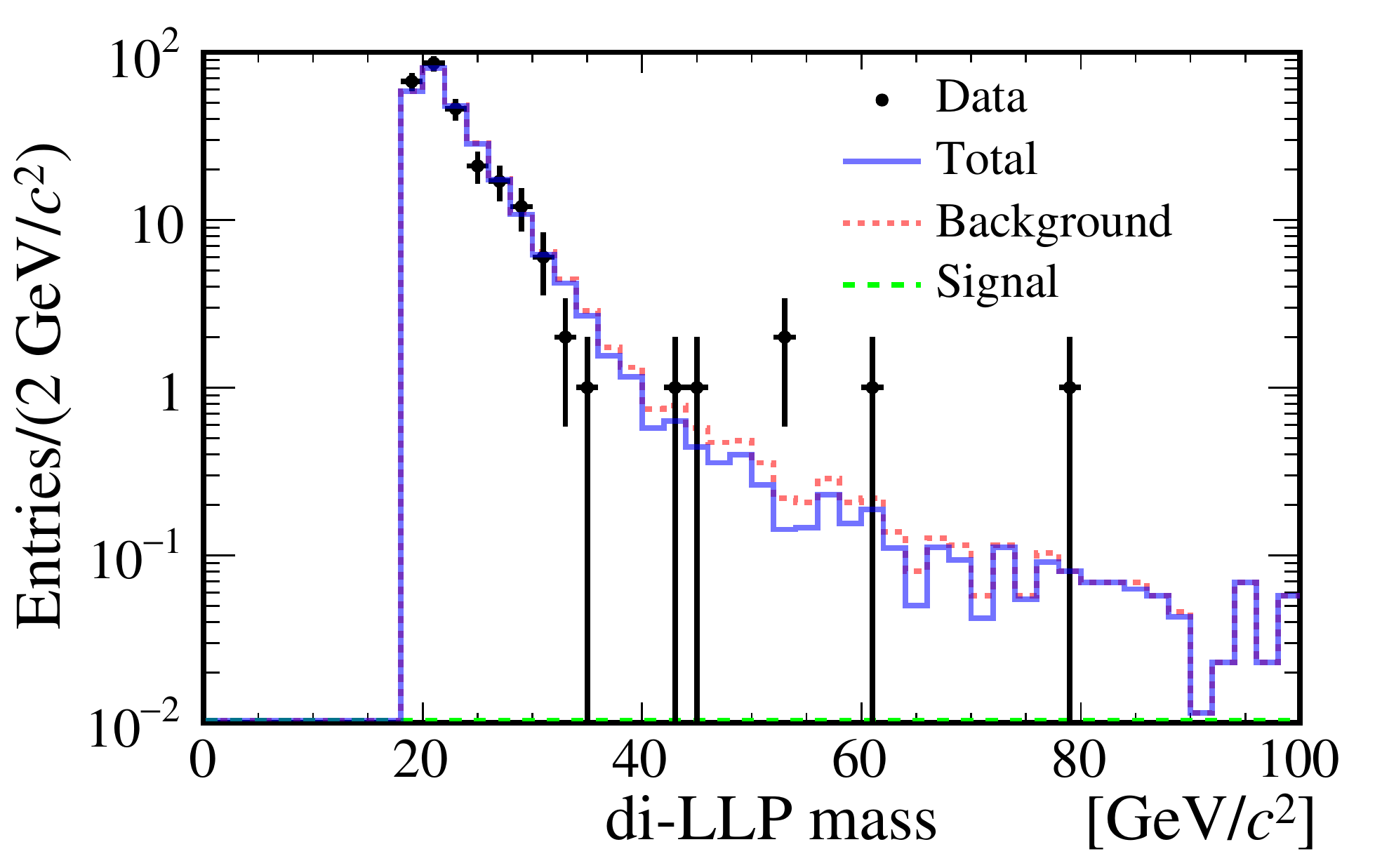}   \put(-160,155){(b)}\put(-58,158){\plhcbs}
\caption{\small
  Results of the fit based on the model BV48 10ps mH114, for different combinations
  of signal and background selections, (a) signal from \ssix and  background from \btwo,
  (b) signal from \sfour and  background from \bone. Dots with error bars are data,
  the dashed (green) line is the fitted signal and the dotted (red) line the background.
  In both cases the fitted signal is negative. The histogram (blue) is the total fitted function.
}
\label{fig:fit-histopdf-bv48-2}
\end{center}
\end{figure}

The signal template is the histogram built from BV simulated events selected under the same conditions as data,
\ie \ssix.
The background template is the histogram obtained from data events selected by the \bone conditions.
The number of signal (background) candidates $N_s$ ($N_b$) is
determined by an extended maximum likelihood fit.
The results are given in Fig.~\ref{fig:fit-histopdf-bv48}
for the BV48 10ps mH114 signal. The fit \chisqndf is 0.6.
Note that only the portion of the di-LLP mass spectrum above 19\gevcc is used,
in order to be sufficiently above the mass threshold set by the selections.
Alternatively, \sfour and \btwo  are used to assess systematic effects.
The fit results for the selections (\ssix,\btwo), (\sfour,\bone) are shown in Fig.~\ref{fig:fit-histopdf-bv48-2}.
The corresponding fit \chisqndf values are 0.6 and 1.0.
The results are given in Table~\ref{tab:fit-histopdf-all} for all fully simulated signal models.
All fits give a negative number of signal candidates, compatible with zero.
These results are correlated because the data sample is in common and
the di-LLP mass shapes are almost identical for the different fully simulated models as depicted in Fig.~\ref{fig:diLLP}.
A check is performed on 142 di-LLP
candidates selected from simulated \bbbar background without the requirement 
on \RXY and with $\mllpmin=4\gevcc$ for both LLPs.
The fitted number of signal events is $-0.8\pm3.5$.

\begin{figure}[!h]
  \begin{center}
    \includegraphics[width=0.64\linewidth]{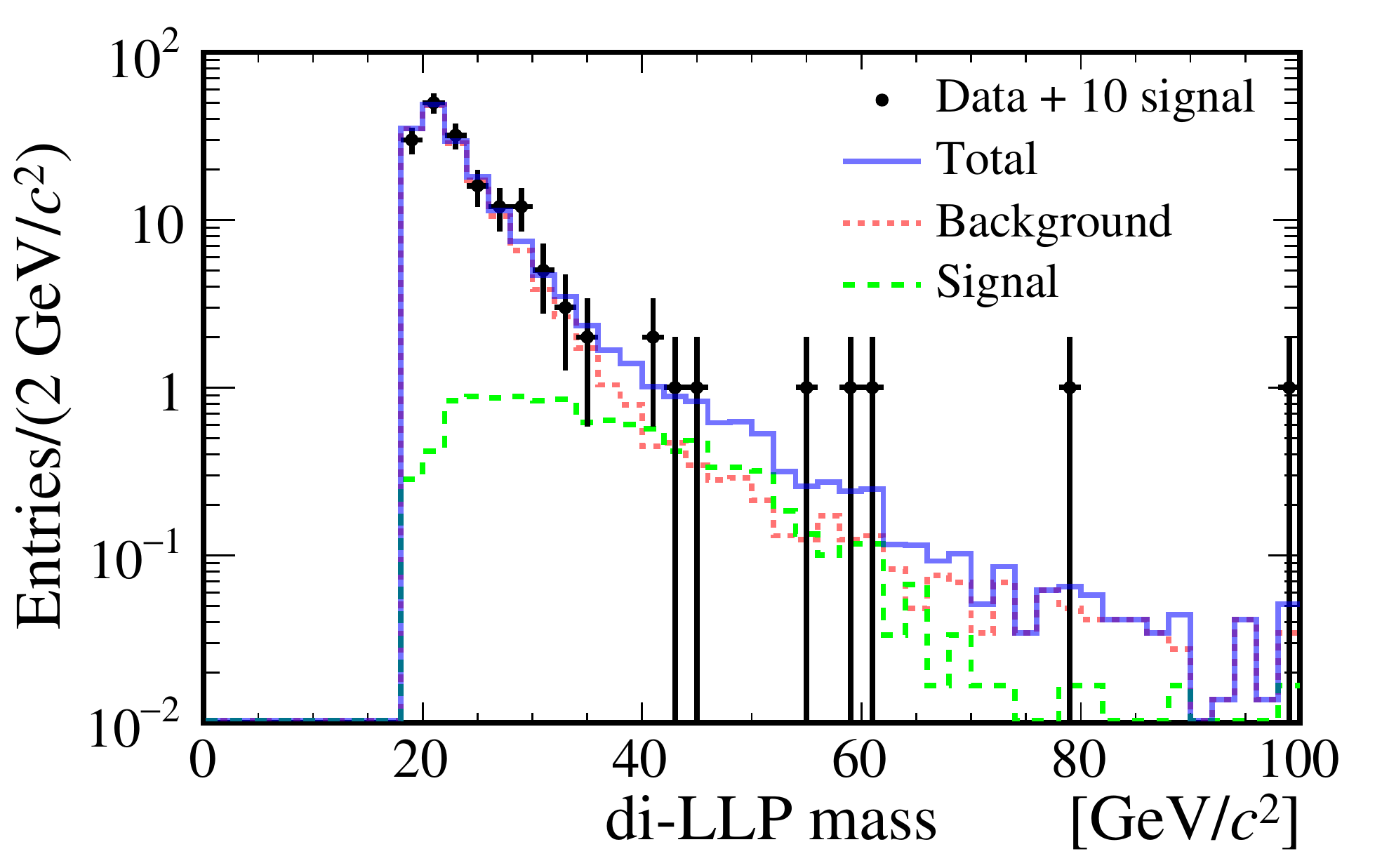} \put(-160,155){(a)}\put(-245,155){\plhcbs} \\
    \includegraphics[width=0.64\linewidth]{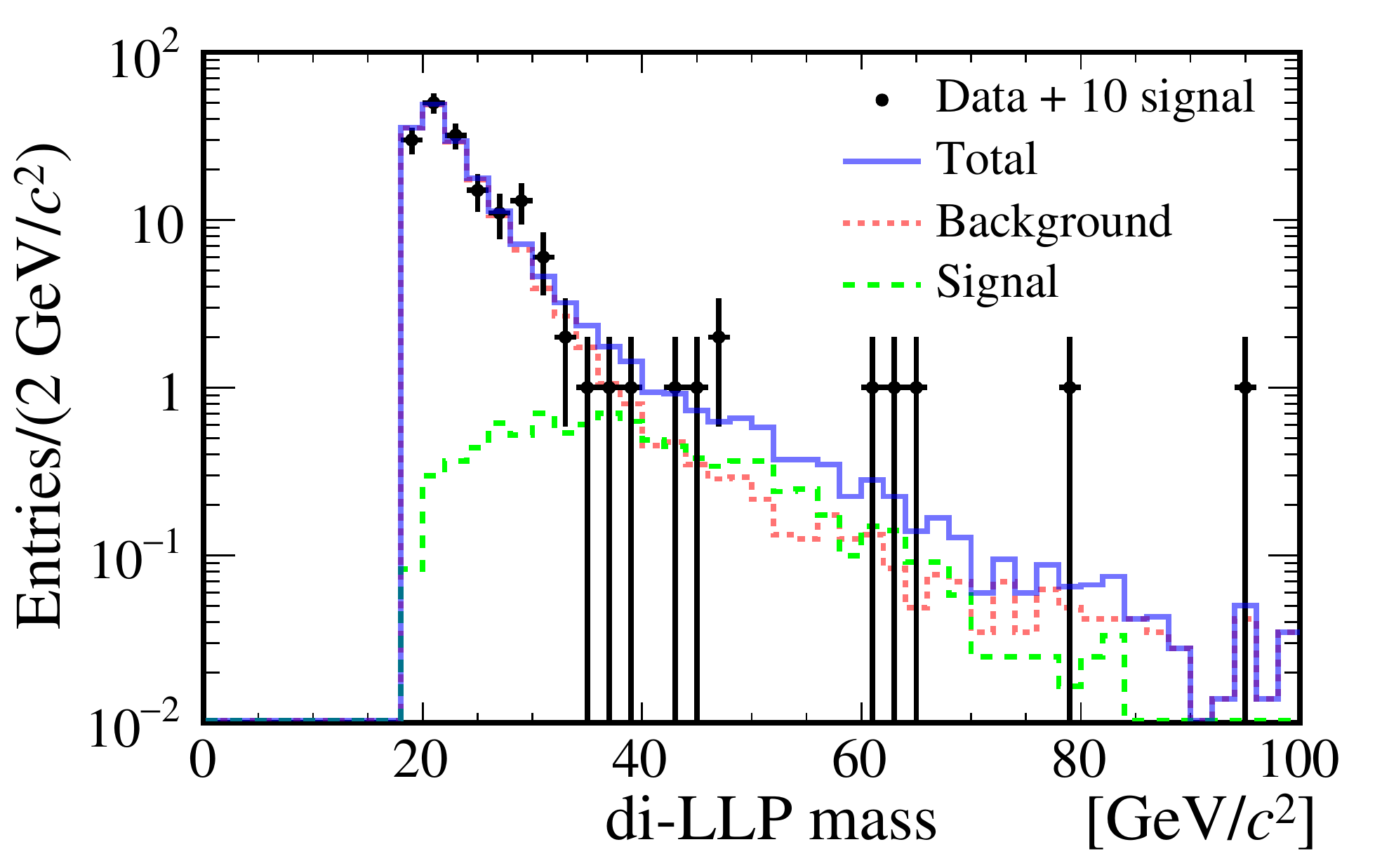} \put(-160,155){(b)}\put(-245,158){\plhcbs} 
\caption{\small
  Results of the fit to the data to which
    10 signal events have been added randomly chosen following the signal model.
    For the theoretical model BV48 10ps mH100, in (a), the fitted signal is $11.1\pm7.0$ events;
    for BV48 10ps mH125, in (b), the result is   $9.3\pm5.6$ events.
}
\label{fig:fit-signaladded}
\end{center}
\end{figure}

The behaviour and sensitivity of the procedure is further studied by adding a small number of signal
events to the data according to a given signal model.
Figure~\ref{fig:fit-signaladded} shows the results for two models with 10 signal events added
to the data. The fitted $N_s$ corresponds well to the number of injected signal events.

\begin{table}[tb]
  \caption{\small
    Values of the fitted signal and background events for the different fully simulated signal models.
    The signal/background combinations are defined in the first row.
}
\begin{center}
\begin{tabular}{lcccc}
\hline
&  \multicolumn{2}{c}{ (\ssix,\bone) }      & (\ssix,\btwo) & (\sfour,\bone) \\
\cline{2-3}
Model     &  $N_s$   & $N_b$ & $N_s$ & $N_s$\\
\hline
BV48 5ps   mH114 &  $-2.6 \pm$  4.4 & 163.6 $\pm 13.6$ &  $-4.8 \pm  3.9$  &  $-1.7\pm 3.9$\\
BV48 10ps  mH114 &  $-3.3 \pm$  3.5 & 164.3 $\pm 13.4$ &  $-4.6 \pm  3.1$  &  $-3.1\pm 3.6$\\
BV48 15ps  mH114 &  $-3.5 \pm$  3.6 & 164.5 $\pm 13.5$ &  $-4.4 \pm  3.1$  &  $-2.0\pm 3.6$\\
BV48 50ps  mH114 &  $-1.4 \pm$  3.6 & 162.4 $\pm 13.3$ &  $-2.7 \pm  3.4$  &  $-2.1\pm 4.2$\\
BV48 100ps mH114 &  $-0.7 \pm$  4.1 & 161.7 $\pm 13.4$ &  $-3.5 \pm  3.9$  &  $-3.2\pm 4.2$\\
BV35 10ps  mH114 &  $-4.3 \pm$  3.3 & 165.3 $\pm 13.4$ &  $-5.9 \pm  3.1$  &  $-4.6\pm 3.5$\\
BV20 10ps  mH114 &  $-1.9 \pm$  1.6 & 162.8 $\pm 12.9$ &  $-2.7 \pm  1.7$  &  $-2.0\pm 2.4$\\
BV48 10ps  mH100 &  $-1.7 \pm$  4.7 & 162.7 $\pm 13.7$ &  $-4.4 \pm  4.4$  &  $-5.2\pm 4.7$\\
BV48 10ps  mH125 &  $-2.8 \pm$  3.5 & 163.8 $\pm 13.4$ &  $-4.1 \pm  3.2$  &  $-3.2\pm 3.6$\\
BV55 10ps  mH114 &  $-3.1 \pm$  3.7 & 164.1 $\pm 13.5$ &  $-4.6 \pm  3.4$  &  $-1.1\pm 3.7$\\
BV55 10ps  mH125 &  $-2.6 \pm$  3.5 & 163.6 $\pm 13.4$ &  $-4.0 \pm  3.2$  &  $-3.9\pm 3.8$\\
\hline
\end{tabular}
\label{tab:fit-histopdf-all}
\end{center}
\end{table}

An alternative fit procedure has been applied, using parameterised signal and background templates.
The sum of two exponential functions is used for the background,
and an exponential convolved with a Gaussian function for the signal.
The results are consistent with a null signal for all the models.

As a final check a two-dimensional sideband subtraction method (``ABCD method''~\cite{ABCD})
has been applied in the reconstructed mass of one LLP and the number of tracks of the other LLP,
also giving results consistent with zero signal.

\section{Detection efficiency and systematic uncertainties}
The determination of the detection efficiency is based on simulated events. The geometrical acceptance
for the detection of one \khi in LHCb is, depending on the model, between 20 and 30\%.
After selection \ssix the predicted total di-LLP detection efficiency is in the range 0.1--1\% for most
of the models.
Potential discrepancies between simulation and data are considered as sources of systematic uncertainties.
Table~\ref{tab:systematics} summarises the contributions of the systematic uncertainties,
which are valid for all fully simulated models, dominated by the 15\% contribution from the trigger.

\begin{table}[htb!]
  \caption{\small Contributions to the systematic uncertainty for fully simulated models.
    For the analysis based on the fast simulation the same total systematic uncertainty is
    adopted augmented by 5\% to account for the relative imprecision of the fast and
    full simulations.
    The contributions from the signal and the data-driven background models used in      
    the di-LLP mass fit are discussed in the text.
    }
\begin{center}
\begin{tabular}{ l c}
\hline
 Source                        &  (\%)     \\
\hline
 Trigger                       &  15      \\
 Track reconstruction          &  5       \\
 Vertex reconstruction         &  4      \\
 \pt\ and mass calibration     &  6        \\     
 Material veto                 &  4        \\
 PV multiplicity               &  0.1      \\
 Beam line position            &  0.7      \\
 Theoretical model             &  9.9        \\
 Integrated luminosity         &  \LLE     \\ 
 \hline
  Total                        &  \SYST     \\
\hline
\end{tabular} 
\label{tab:systematics}
\end{center}
\end{table}

The consistency between the trigger efficiency in data and simulation is     
checked by selecting LLP events with an independent trigger, designed for     
the detection of \jpsi events. Comparing the fraction of the data that        
also passes the double-LLP selection with the corresponding fraction in       
simulated inclusive \jpsi events, consistent efficiencies are found within    
a statistical uncertainty of 30\%. A more precise result is obtained when     
requiring only a single LLP candidate~\cite{LHCb-PAPER-2014-062} and          
assuming uncorrelated contributions from the two LLPs to determine the        
efficiency for detecting two LLPs in coincidence. A maximum discrepancy       
between data and simulation of 15\% is inferred, which is the value adopted.
                                                                                
  The consistency between the track reconstruction efficiency in data and simulation is studied
  by a comparison of the number of tracks selected in displaced vertices from \bbbar events.
  The average number of tracks per LLP in data is higher than in simulated events by about 0.07 tracks.
  Assuming that this small effect is entirely due to a difference in tracking efficiency,
  the overall di-LLP detection efficiency changes by at most 5\%.

  The vertex reconstruction efficiency is affected by the tracking efficiency and
  resolution. A study of vertices from
  $B^0 \rightarrow \jpsi \Kstarz$ with $\decay{\jpsi}{\mumu}$ and $\Kstarz \rightarrow K^+ \pi^-$
  has shown that the data and simulation detection efficiencies
  for this four-prong process agree within 7.5\%~\cite{LHCb-PAPER-2014-062}.
  This has been evaluated
  to correspond at most to a 4\% discrepancy between the di-LLP efficiency in data and simulation.

  A maximum mismatch of 10\% on both the transverse momentum and mass scales is inferred from
  the comparison of data and simulated \bbbar distributions, which propagates to a
  6\% contribution to the systematic uncertainty.

  The effect of the material veto corresponds to a reduction of the geometrical acceptance and
  depends mainly on the LLP lifetime.
  An analysis with the requirement of $\RXY< 4\mm$ allows to infer a maximum systematic uncertainty of 4\%.

  A small contribution to the systematic uncertainty of 0.1\%
  is determined by reweighting the simulated events to match the PV multiplicity in the data.

  The uncertainty on the position of the beam line is less than 20\mum~\cite{LHCb-DP-2014-001}.
  It can affect the secondary vertex selection, mainly via the requirement on \RXY.
  By altering the PV position in simulated signal events, the maximum effect on the di-LLP selection efficiency
  is 0.7\%.

  The Higgs-like particle production model is mainly affected by the uncertainty on the parton
  luminosity. 
  A maximum variation of the detection efficiency of 9.5\% is obtained following
  the prescriptions given in~\cite{PDF4LHC}.
  A second contribution of 3\% is obtained by reweighting the \pythia generated events to
  match a recent calculation of the \pt distributions~\cite{higgs-pt}. 
  The total theoretical uncertainty is 9.9\%, obtained    by summing in quadrature
  the mentioned contributions.

In addition to the systematic uncertainty on the detection efficiency, the following contributions
have been considered.
The uncertainty on the integrated luminosity is 1.7\%~\cite{LHCb-PAPER-2014-047}.
As previously stated, the uncertainty on the momentum scale and the invariant mass scale is
smaller than 10\%.
This value is also assumed for the di-LLP mass calibration.
To assess the impact on the signal measurement, pseudoexperiments are produced 
with 10 events of simulated signal added to the background following the nominal
signal distribution but with the di-LLP mass value scaled by $\pm10\%$.
The subsequent maximum variation of the fitted number of events is $\pm 1.6$, for all the signal hypotheses.
The uncertainty due to the shape of the background template
is obtained by comparing the number of fitted events obtained with the \bone and \btwo selections.
The change is less than one event, for all the signal models.
The difference in data and simulation in the  di-LLP mass resolution and
the statistical precision of the signal templates used in the fit have a negligible effect.
Hence, a fit uncertainty of $\pm$2 events is considered in the calculation
of the cross-section upper limits.

For the analysis based on the fast simulation, a 5\% uncertainty is added 
to account for the relative imprecision of the fast simulation with respect 
to the full simulation, as explained in Section~\ref{sec:evtgen}.

\section{Results}

The 95\% confidence level (CL) upper limits on the production cross-section times branching ratio
are presented in Table~\ref{tab:UL1}, for the fully simulated models, based on the CLs approach~\cite{art:cls}.
\begin{table}[b!]
\caption{\small
  Detection efficiency with total uncertainty,
  and upper limits at 95\% CL on the cross-section times branching ratio
  for the process $ pp \rightarrow \ho X$, $\ho \rightarrow\khi\khi \rightarrow 6q$
  for the fully simulated models.
}
\renewcommand{\arraystretch}{1.3}
\begin{center}
 \begin{tabular}{lcccc}
 \hline
 Model      & Efficiency   & Expected      & Observed       \\[-1mm]
 &              [\%]       &  upper limit  [\pb] & upper limit  [\pb]  \\
\hline
BV48 5ps mH114   &  0.528 $\pm$  0.114 &     3.2$^{+ 2.1}_{- 1.1}$ &     3.5 \\
BV48 10ps mH114  &  0.925 $\pm$  0.194 &     1.8$^{+ 1.2}_{- 0.6}$ &     1.7 \\
BV48 15ps mH114  &  0.966 $\pm$  0.208 &     1.8$^{+ 1.2}_{- 0.6}$ &     1.6 \\
BV48 50ps mH114  &  0.419 $\pm$  0.090 &     4.6$^{+ 2.9}_{- 1.6}$ &     4.4 \\
BV48 100ps mH114 &  0.171 $\pm$  0.037 &  11.9$^{+  7.1}_{-  3.8}$ &     12.3 \\
BV35 10ps mH114  &  0.268 $\pm$  0.058 &     5.6$^{+ 3.8}_{- 2.0}$ &     4.9 \\
BV20 10ps mH114  &  0.016 $\pm$  0.003 &     52$^{+ 38}_{- 20}$    &     54 \\
BV48 10ps mH100  &  0.864 $\pm$  0.186 &     2.5$^{+ 1.6}_{- 0.8}$ &     2.6 \\
BV48 10ps mH125  &  0.771 $\pm$  0.166 &     2.0$^{+ 1.4}_{- 0.7}$ &     2.0 \\
BV55 10ps mH114  &  0.851 $\pm$  0.183 &     1.9$^{+ 1.3}_{- 0.7}$ &     1.9 \\
BV55 10ps mH125  &  0.937 $\pm$  0.201 &     1.7$^{+ 1.1}_{- 0.6}$ &     1.7 \\
\hline
\end{tabular}
\label{tab:UL1}
\end{center}
\end{table}
The fast simulation allows the exploration of a larger region of parameter space.
The cross-section times branching fraction upper limits at 95\% CL
for benchmark theoretical models are shown in Fig.~\ref{fig:fast-fit-CLs} (the corresponding
tables are given in Appendix~\ref{app:ULfast}).

The estimated detection efficiencies
can be found in Appendix~\ref{app:eff}, Tables~\ref{tab:fast-eps-m} and~\ref{tab:fast-eps-tau}.
The efficiency increases with \mllp because more particles are produced in the decay of heavier LLPs.
This effect is only partially counteracted by the
loss of particles outside of the spectrometer acceptance,
which is especially the case with heavier Higgs-like particles.
Another competing phenomenon is that
the lower boost of heavier LLPs results in a shorter average flight length, \ie
the requirement of a minimum \RXY disfavours heavy LLPs.
The cut on  \RXY is more efficient at selecting LLPs with large lifetimes, but 
for lifetimes larger than $\sim50\ps$
a portion of the decays falls into the material region and is discarded.
Finally, a drop of sensitivity is expected for LLPs with a
lifetime close to the \bquark hadron lifetimes,
where the contamination from \bbbar events becomes important, especially for low mass LLPs.

\begin{figure}[hbt!]
\centering
\includegraphics[height=5cm, angle=0]{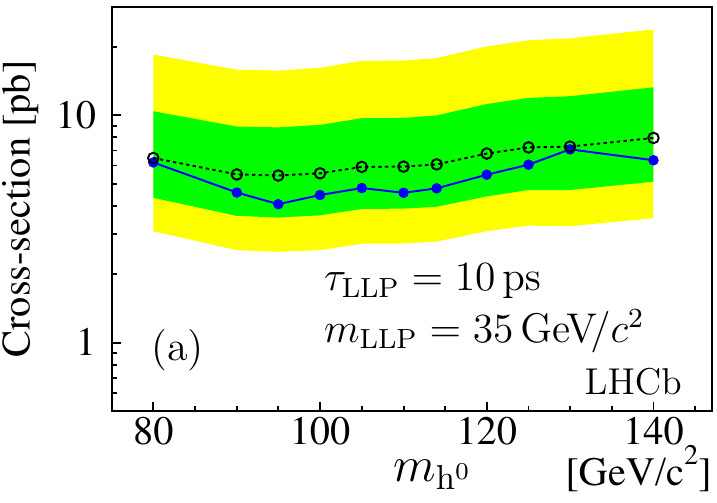} 
\includegraphics[height=5cm, angle=0]{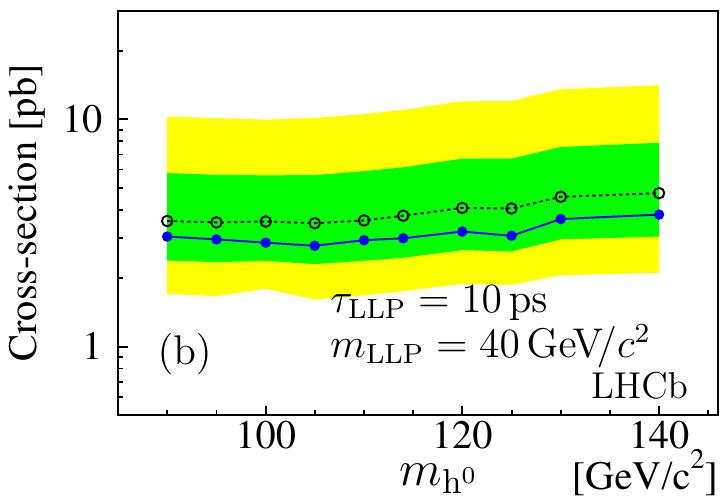} 
 \includegraphics[height=5cm, angle=0]{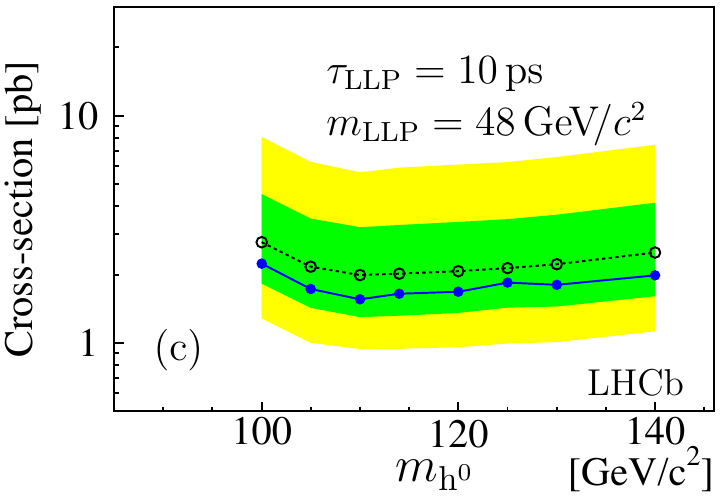} 
 \includegraphics[height=5cm, angle=0]{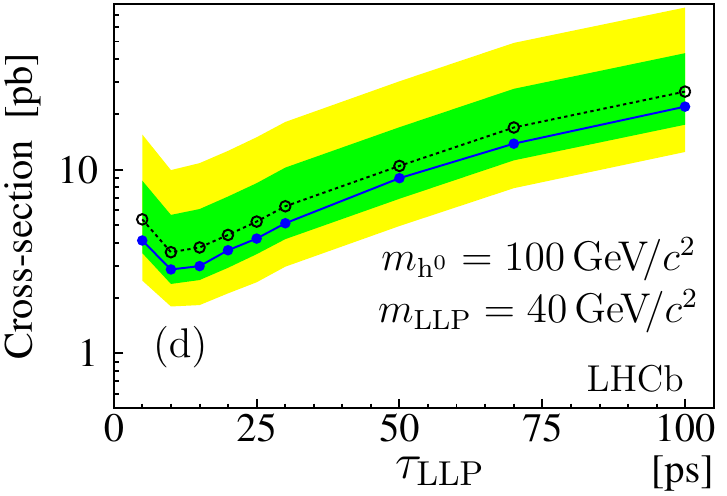} 
 \includegraphics[height=5.0cm, angle=0]{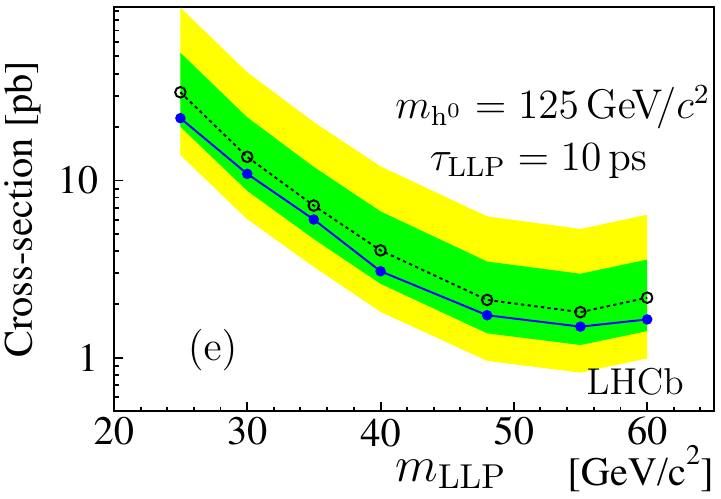} 
\includegraphics[height=5cm, angle=0]{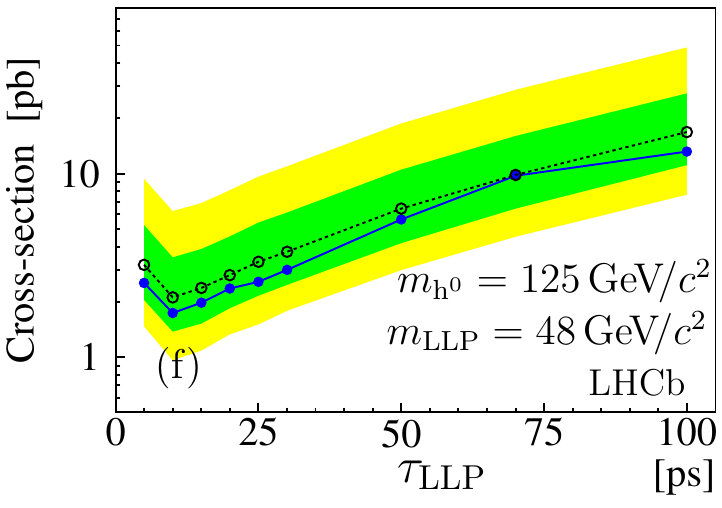} 
\caption{\small
Expected (open dots with 1$\sigma$ and 2$\sigma$ bands) and observed (full dots) upper limits at
95\% confidence level, (a) -- (c) shown for different masses of the Higgs-like particle, (d) and (f) for
different LLP lifetimes, and (e) as a function of the LLP mass. The values of the other
parameters are indicated on the plots. Results inferred from the fast simulation.
}
\label{fig:fast-fit-CLs}
\end{figure}

\section{Conclusion}

A search for Higgs-like bosons decaying into
two long-lived particles decaying hadronically has been carried out
using data from $pp$ collisions at 7\tev collected with the LHCb detector,
corresponding to a total integrated luminosity of \LL\invfb.

The model used to describe the LLP decay is an mSUGRA process
in which the lightest neutralino $\khi$ decays
through a baryon number violating coupling to three quarks.
Upper limits have been placed on the production cross-section for
Higgs-like boson masses from 80 to 140\gevcc,
LLP masses in the range 20--60\gevcc, and LLP lifetimes in the range of 5--100\ps.
The number of candidates is determined by the di-LLP invariant mass fit with
signal templates
inferred from simulation, and  background estimates from data.
For the explored parameter space of the theory all results, which are correlated, are consistent with zero.
Upper limits at 95\% CL for cross-section times branching ratio of 1 to 5\pb  are inferred for most
of the considered parameter range. They are below 2\pb for
the decay of a 125\gevcc Higgs-like particle in two LLPs with mass in the 48--60\gevcc range and
10\ps lifetime.

\clearpage

\clearpage
\section*{Acknowledgements}
\noindent We express our gratitude to our colleagues in the CERN
accelerator departments for the excellent performance of the LHC. We
thank the technical and administrative staff at the LHCb
institutes. We acknowledge support from CERN and from the national
agencies: CAPES, CNPq, FAPERJ and FINEP (Brazil); NSFC (China);
CNRS/IN2P3 (France); BMBF, DFG and MPG (Germany); INFN (Italy); 
FOM and NWO (The Netherlands); MNiSW and NCN (Poland); MEN/IFA (Romania); 
MinES and FANO (Russia); MinECo (Spain); SNSF and SER (Switzerland); 
NASU (Ukraine); STFC (United Kingdom); NSF (USA).
We acknowledge the computing resources that are provided by CERN, IN2P3 (France), KIT and DESY (Germany), INFN (Italy), SURF (The Netherlands), PIC (Spain), GridPP (United Kingdom), RRCKI and Yandex LLC (Russia), CSCS (Switzerland), IFIN-HH (Romania), CBPF (Brazil), PL-GRID (Poland) and OSC (USA). We are indebted to the communities behind the multiple open 
source software packages on which we depend.
Individual groups or members have received support from AvH Foundation (Germany),
EPLANET, Marie Sk\l{}odowska-Curie Actions and ERC (European Union), 
Conseil G\'{e}n\'{e}ral de Haute-Savoie, Labex ENIGMASS and OCEVU, 
R\'{e}gion Auvergne (France), RFBR and Yandex LLC (Russia), GVA, XuntaGal and GENCAT (Spain), Herchel Smith Fund, The Royal Society, Royal Commission for the Exhibition of 1851 and the Leverhulme Trust (United Kingdom).

\clearpage
{\noindent\normalfont\bfseries\Large Appendices}

\appendix
\section{Fully simulated signal datasets}\label{app:gen}

Table~\ref{tab:models-full} shows the parameters used to generate the 11
fully simulated signal models with \pythia~6.
The Higgs-like boson is produced by gluon-gluon fusion.
In the table M$_1$ corresponds to the \pythia parameter RMSS(1), and \tanb \, to RMSS(5).
In addition, M$_2$ (RMSS(2)) is set at 250\gevcc and $\mu$ (RMSS(4)) has the value 140.
A \mhzero value of 125\gevcc requires RMSS(16) = 2300.

\label{sec:full-gen}
\begin{table}[htb!]
\caption{\small
Parameters of the signal models generated by \pythia and fully simulated.
}
\begin{center}
\begin{tabular}{l c c c c c }
\hline
Model             & M$_1$       & \tanb   & $\mhzero$  & $\mllp$   & $\taullp$  \\
                  & $[\gevcc]$  &         &  [\gevcc]  & [\gevcc]  &  [\ps]  \\
\hline
BV48 5ps mH114    & 62          & 5       & 114  & 48  &  5        \\
BV48 10ps mH114   & 62          & 5       & 114  & 48  & 10        \\
BV48 15ps mH114   & 62          & 5       & 114  & 48  & 15        \\
BV48 50ps mH114   & 62          & 5       & 114  & 48  & 50        \\	
BV48 100ps mH114  & 62          & 5       & 114  & 48  & 100       \\	
BV35 10ps mH114   & 46          & 5       & 114  & 35  & 10        \\
BV20 10ps mH114   & 28          & 5       & 114  & 20  & 10        \\
BV48 10ps mH100   & 71          & 2.4     & 100  & 48  & 10        \\
BV48 10ps  mH125  & 60          & 8       & 125  & 48  & 10      \\
BV55 10ps  mH114  & 71          & 5.1     & 114  & 55  & 10      \\
BV55 10ps  mH125  & 69          & 6.2     & 125  & 55  & 10      \\
\hline
\end{tabular} 
\label{tab:models-full}
\end{center}
\end{table}

\clearpage

\section{Detection efficiencies}\label{app:eff}

Table~\ref{tab:fast-eps-m} gives the detection efficiency as a function of \mhzero and \mllp,
the LLP lifetime is 10\ps.
Table~\ref{tab:fast-eps-tau} gives the efficiency as a function of \mllp and \taullp, assuming $\mhzero=114\gevcc$.

\begin{table}[b!]
\caption{\small
  Detection efficiency values in percent estimated by the fast simulation as a function of  \mhzero and  \mllp .
  The LLP lifetime is 10 ps. The statistical uncertainty is
10\% for $\epsilon \sim 0.02\%$,  
5 \% for $\epsilon \sim 0.1\%$,
3\% for $\epsilon \sim 0.5\%$, and 
2\% for $\epsilon \sim 1\%$.
}
\begin{center}
\scalebox{0.94}{
\begin{tabular}{c c c c c c c c c}
\hline
   \mhzero             &  \multicolumn{7}{c}{\mllp [GeV/c$^2$]} \\
$\rm [GeV/c^2]$              &     20   &     25   &     30   &     35   &     40   &     48   &     55   & 60  \\
\hline
  80                    &    0.035 &    0.126 &    0.276 &    0.514 &    \text{--} &    \text{--} &    \text{--} &    \text{--} \\
  90                    &    0.027 &    0.084 &    0.213 &    0.456 &    0.699 &    \text{--} &    \text{--} &    \text{--} \\
  95                    &    0.023 &    0.077 &    0.203 &    0.414 &    0.689 &    \text{--} &    \text{--} &    \text{--} \\
 100                    &    0.025 &    0.073 &    0.184 &    0.368 &    0.647 &    0.858 &    \text{--} &    \text{--} \\
 105                    &    0.018 &    0.066 &    0.139 &    0.324 &    0.574 &    1.018 &    \text{--} &    \text{--} \\
 110                    &    0.017 &    0.053 &    0.146 &    0.291 &    0.525 &    1.016 &    \text{--} &    \text{--} \\
 114                    &    0.014 &    0.048 &    0.134 &    0.259 &    0.472 &    0.963 &    0.817 &    \text{--} \\
 120                    &    0.016 &    0.047 &    0.107 &    0.222 &    0.402 &    0.836 &    1.013 &    \text{--} \\
 125                    &    0.009 &    0.042 &    0.097 &    0.225 &    0.377 &    0.765 &    0.997 &    \text{--} \\
 130                    &    0.014 &    0.037 &    0.085 &    0.191 &    0.325 &    0.708 &    0.914 &    0.991 \\
 140                    &    0.002 &    0.031 &    0.075 &    0.163 &    0.277 &    0.566 &    0.782 &    0.881 \\
 \hline
\end{tabular} 
\label{tab:fast-eps-m}
}
\end{center}
\end{table}

\begin{table}[!]
\caption{\small
Detection efficiency in percent estimated by the fast simulation as a function of the \mllp and \taullp,
for \mhzero=114 GeV/c$^2$.
The statistical uncertainty is
10\% for $\epsilon \sim 0.02\%$,  
5 \% for $\epsilon \sim 0.1\%$,
3\% for $\epsilon \sim 0.5\%$, and 
2\% for $\epsilon \sim 1\%$.
} 
\begin{center}
  \scalebox{0.94}{
\begin{tabular}{c c c c c c c c }
  \hline
  \taullp         &  \multicolumn{7}{c}{ \mllp  [GeV/c$^2$]}\\
   $\rm [ps]$          &     20   &     25   &     30   &     35   &     40   &     48   &     55   \\
\hline
         5                 &    0.021 &    0.053 &    0.129 &    0.234 &    0.366 &    0.545 &    0.289 \\
        10                 &    0.014 &    0.048 &    0.134 &    0.259 &    0.472 &    0.963 &    0.817 \\
        15                 &    0.013 &    0.042 &    0.113 &    0.198 &    0.389 &    0.932 &    1.052 \\
        20                 &    0.007 &    0.035 &    0.083 &    0.174 &    0.338 &    0.834 &    1.150 \\
        25                 &    0.006 &    0.034 &    0.073 &    0.148 &    0.289 &    0.731 &    1.126 \\
        30                 &    0.005 &    0.026 &    0.066 &    0.128 &    0.241 &    0.643 &    1.091 \\
        40                 &    0.003 &    0.017 &    0.044 &    0.114 &    0.193 &    0.490 &    0.960 \\
        50                 &    0.004 &    0.015 &    0.035 &    0.082 &    0.157 &    0.397 &    0.806 \\
        70                 &    0.002 &    0.009 &    0.021 &    0.062 &    0.104 &    0.280 &    0.596 \\
       100                 &    0.001 &    0.005 &    0.015 &    0.033 &    0.071 &    0.178 &    0.383 \\
\hline
\end{tabular} 
\label{tab:fast-eps-tau}
}
\end{center}
\end{table}

\section{Cross-section upper limits tables}\label{app:ULfast}

Expected and observed  95\% CL cross-section times branching ratio upper limits for
benchmark models, from the fast simulation.
Table~\ref{tab:fast-mass-CLs-1} and \ref{tab:fast-mass-CLs-2} give the limits as a function of $\mhzero$,
covering LLP masses from 35 to 60\gevcc, $\taullp=10\ps$.
Table~\ref{tab:fast-time-CLs}: limits as a function of the LLP lifetime for $\mhzero=100\gevcc$ and
$\mllp=40\gevcc$, and for $\mhzero=125\gevcc$ and $\mllp=48\gevcc$.
Table~\ref{tab:fast-125-CLs}:  limits
as a function of the LLP mass, for  $\mhzero=125\gevcc$, $\taullp=10\ps$.
\begin{table}[h!]
  \caption{ \small Expected and observed
    95\% CL cross-section  times branching ratio upper limits as a function of \mhzero,
    with $\mllp=35\gevcc$, and $\taullp=10\ps$, estimated by the fast simulation.
  }\label{tab:fast-mass-CLs-1}
\begin{center}
\scalebox{1.}{
\renewcommand{\arraystretch}{1.3}
 \begin{tabular}{lccc}
 \hline
 Model     & Expected      & Observed       \\
 & upper limit [pb]   & upper limit [pb]   \\
 \hline
BV35 10ps mH80   &     6.49$^{+ 3.94}_{- 2.16}$ &     6.20 \\
BV35 10ps mH90   &     5.50$^{+ 3.42}_{- 1.89}$ &     4.56 \\
BV35 10ps mH95   &     5.42$^{+ 3.41}_{- 1.88}$ &     4.06 \\
BV35 10ps mH100  &     5.55$^{+ 3.52}_{- 1.92}$ &     4.45 \\
BV35 10ps mH105  &     5.92$^{+ 3.79}_{- 2.06}$ &     4.78 \\
BV35 10ps mH110  &     5.94$^{+ 3.79}_{- 2.06}$ &     4.56 \\
BV35 10ps mH114  &     6.07$^{+ 3.92}_{- 2.11}$ &     4.77 \\
BV35 10ps mH120  &     6.79$^{+ 4.42}_{- 2.39}$ &     5.47 \\
BV35 10ps mH125  &     7.21$^{+ 4.70}_{- 2.54}$ &     6.03 \\
BV35 10ps mH130  &     7.28$^{+ 4.83}_{- 2.59}$ &     7.08 \\
BV35 10ps mH140  &     7.95$^{+ 5.32}_{- 2.85}$ &     6.35 \\
\hline
\end{tabular}
}
\end{center}
\end{table}
 
\begin{table}[b]
  \caption{ \small Expected and observed
  95\% CL cross-section times branching ratio upper limits as a function of
  \mhzero, for LLP masses of 40, 48, 55, and 60\gevcc, $\taullp=10\ps$, estimated by the fast simulation.
  }\label{tab:fast-mass-CLs-2}
\begin{center}
\scalebox{1.}{
\renewcommand{\arraystretch}{1.3}
 \begin{tabular}{lccc}
 \hline
 Model     & Expected      & Observed       \\
 & upper limit [pb]   & upper limit [pb]   \\
 \hline

BV40 10ps mH90   &     3.57$^{+ 2.23}_{- 1.18}$ &     3.04 \\
BV40 10ps mH95   &     3.52$^{+ 2.18}_{- 1.17}$ &     2.96 \\
BV40 10ps mH100  &     3.55$^{+ 2.12}_{- 1.16}$ &     2.86 \\
BV40 10ps mH105  &     3.49$^{+ 2.19}_{- 1.18}$ &     2.77 \\
BV40 10ps mH110  &     3.59$^{+ 2.32}_{- 1.21}$ &     2.93 \\
BV40 10ps mH114  &     3.76$^{+ 2.38}_{- 1.30}$ &     2.99 \\
BV40 10ps mH120  &     4.07$^{+ 2.63}_{- 1.42}$ &     3.20 \\
BV40 10ps mH125  &     4.04$^{+ 2.66}_{- 1.43}$ &     3.07 \\
BV40 10ps mH130  &     4.55$^{+ 2.98}_{- 1.61}$ &     3.63 \\
BV40 10ps mH140  &     4.71$^{+ 3.14}_{- 1.69}$ &     3.79 \\
  \hline
BV48 10ps mH100  &     2.78$^{+ 1.75}_{- 0.95}$ &     2.23 \\
BV48 10ps mH105  &     2.17$^{+ 1.36}_{- 0.74}$ &     1.73 \\
BV48 10ps mH110  &     1.99$^{+ 1.24}_{- 0.69}$ &     1.56 \\
BV48 10ps mH114  &     2.02$^{+ 1.29}_{- 0.70}$ &     1.65 \\
BV48 10ps mH120  &     2.07$^{+ 1.34}_{- 0.71}$ &     1.68 \\
BV48 10ps mH125  &     2.12$^{+ 1.38}_{- 0.74}$ &     1.74 \\
BV48 10ps mH130  &     2.22$^{+ 1.45}_{- 0.78}$ &     1.80 \\
BV48 10ps mH140  &     2.49$^{+ 1.65}_{- 0.89}$ &     1.98 \\
\hline
BV55 10ps mH130 &     1.94$^{+ 1.27}_{- 0.69}$ &     1.76 \\
BV55 10ps mH140 &     1.93$^{+ 1.26}_{- 0.69}$ &     1.75 \\
\hline
BV60 10ps mH130 &     1.79$^{+ 1.16}_{- 0.63}$ &     1.52 \\
BV60 10ps mH140 &     1.86$^{+ 1.21}_{- 0.66}$ &     1.48 \\
\hline
\end{tabular}
}
\end{center}
\end{table}

 \begin{table}
  \caption{ \small  Expected and observed 95\% CL cross-section times branching ratio upper limits
    as a function of the LLP lifetime,
    for $\mhzero = 100\gevcc$ and $\mllp=40\gevcc$,
    and for $\mhzero = 125\gevcc$ and $\mllp=48\gevcc$, estimated by the fast simulation.
  }\label{tab:fast-time-CLs}
  \renewcommand{\arraystretch}{1.3}
 \begin{center}
 \begin{tabular}{lccc}
 \hline
 Model     & Expected      & Observed       \\
           & upper limit [pb]  & upper limit  [pb]   \\
\hline
BV40  5ps mH100  &     5.36$^{+ 3.36}_{- 1.85}$ &     4.11 \\
BV40 10ps mH100  &     3.55$^{+ 2.12}_{- 1.16}$ &     2.86 \\
BV40 15ps mH100  &     3.76$^{+ 2.34}_{- 1.26}$ &     2.98 \\
BV40 20ps mH100  &     4.41$^{+ 2.73}_{- 1.49}$ &     3.63 \\
BV40 25ps mH100  &     5.21$^{+ 3.23}_{- 1.75}$ &     4.20 \\
BV40 30ps mH100  &     6.32$^{+ 3.95}_{- 2.13}$ &     5.10 \\
BV40 50ps mH100  &     10.5$^{+  6.5}_{-  3.6}$ &      9.0 \\
BV40 70ps mH100  &     17.0$^{+ 10.6}_{-  5.8}$ &     13.8 \\
BV40 100ps mH100 &     26.7$^{+ 16.5}_{-  9.1}$ &     22.1 \\
 \hline
BV48  5ps mH125  &     3.19$^{+ 2.06}_{- 1.14}$ &     2.54 \\
BV48 10ps mH125  &     2.12$^{+ 1.38}_{- 0.74}$ &     1.74 \\
BV48 15ps mH125  &     2.38$^{+ 1.50}_{- 0.86}$ &     1.98 \\
BV48 20ps mH125  &     2.80$^{+ 1.76}_{- 0.95}$ &     2.37 \\
BV48 25ps mH125  &     3.31$^{+ 2.11}_{- 1.15}$ &     2.57 \\
BV48 30ps mH125  &     3.76$^{+ 2.38}_{- 1.28}$ &     2.99 \\
BV48 50ps mH125  &     6.45$^{+ 4.09}_{- 2.26}$ &     5.63 \\
BV48 70ps mH125  &     9.86$^{+ 6.23}_{- 3.42}$ &     9.74 \\
BV48 100ps mH125 &     16.9$^{+ 10.6}_{-  5.8}$ &     13.2 \\
 \hline
 \end{tabular}
 \end{center}
 \end{table}

 \begin{table}
  \caption{ \small  Expected and observed 95\% CL cross-section times branching ratio upper limits
    as a function of the LLP mass, with $\mhzero=125\gevcc$ and $\taullp=10\ps$, estimated by the fast simulation.
  }\label{tab:fast-125-CLs}
  \renewcommand{\arraystretch}{1.3}
 \begin{center}
 \begin{tabular}{lccc}
 \hline
 Model     & Expected      & Observed       \\
           & upper limit  [pb]  & upper limit  [pb]   \\

 \hline
BV20 10ps mH125 &     95.3$^{+ 64.9}_{- 34.7}$ &    112.6 \\
BV25 10ps mH125 &     31.4$^{+ 21.0}_{- 11.3}$ &     22.5 \\
BV30 10ps mH125 &     13.6$^{+  9.1}_{-  4.9}$ &     10.9 \\
BV35 10ps mH125 &     7.21$^{+ 4.70}_{- 2.54}$ &     6.03 \\
BV40 10ps mH125 &     4.04$^{+ 2.66}_{- 1.43}$ &     3.07 \\
BV48 10ps mH125 &     2.12$^{+ 1.38}_{- 0.74}$ &     1.74 \\
BV55 10ps mH125 &     1.81$^{+ 1.17}_{- 0.63}$ &     1.50 \\
BV60 10ps mH125 &     2.18$^{+ 1.40}_{- 0.76}$ &     1.64 \\
\hline
 \end{tabular}
 \end{center}
 \end{table}

\clearpage



\addcontentsline{toc}{section}{References}
\setboolean{inbibliography}{true}
\bibliographystyle{LHCb}
\bibliography{main,LHCb-PAPER,LHCb-CONF,LHCb-DP,LHCb-TDR}

\newpage
\centerline{\large\bf LHCb collaboration}
\begin{flushleft}
\small
R.~Aaij$^{39}$,
B.~Adeva$^{38}$,
M.~Adinolfi$^{47}$,
Z.~Ajaltouni$^{5}$,
S.~Akar$^{6}$,
J.~Albrecht$^{10}$,
F.~Alessio$^{39}$,
M.~Alexander$^{52}$,
S.~Ali$^{42}$,
G.~Alkhazov$^{31}$
% ...

{\footnotesize \it
% ...
$ ^{5}$Clermont Universit{\'e}, Universit{\'e} Blaise Pascal, CNRS/IN2P3, LPC, Clermont-Ferrand, France\\
$ ^{6}$CPPM, Aix-Marseille Universit{\'e}, CNRS/IN2P3, Marseille, France\\
% ...
$ ^{10}$Fakult{\"a}t Physik, Technische Universit{\"a}t Dortmund, Dortmund, Germany\\
% ...
$ ^{31}$Petersburg Nuclear Physics Institute (PNPI), Gatchina, Russia\\
% ...
$ ^{38}$Universidad de Santiago de Compostela, Santiago de Compostela, Spain\\
$ ^{39}$European Organization for Nuclear Research (CERN), Geneva, Switzerland\\
% ...
$ ^{42}$Nikhef National Institute for Subatomic Physics, Amsterdam, The Netherlands\\
% ...
$ ^{47}$H.H. Wills Physics Laboratory, University of Bristol, Bristol, United Kingdom\\
% ...
$ ^{52}$School of Physics and Astronomy, University of Glasgow, Glasgow, United Kingdom\\
% ...
}
\end{flushleft}

%

\end{document}